\documentclass[11pt,twoside,nohyper,color]{JHEP1}

\usepackage{epsfig}
\usepackage{amssymb}
\usepackage{latexsym}
\usepackage{times}
\usepackage{slashed,cite}
\usepackage{upgreek}
\usepackage{rotating}
\usepackage{caption}
\usepackage{amsmath}
\usepackage{amsfonts}
\usepackage{amsbsy}
\usepackage{amscd}
\usepackage{bbm}
\usepackage{graphicx}   
\usepackage{subcaption} 
\usepackage{comment}

\newcommand{\eec}{\end{center}}
\newcommand{\bec}{\begin{center}}

\newcommand{\eem}{\end{matrix}}
\newcommand{\bem}{\begin{matrix}}
\newcommand{\eeq}{\end{equation}}
\newcommand{\beq}{\begin{equation}}
\newcommand{\ba}{\begin{array}}
\newcommand{\ea}{\end{array}}
\newcommand{\bea}{\begin{eqnarray}}
\newcommand{\eea}{\end{eqnarray}}
\newcommand{\baq}{\begin{eqnarray}}
\newcommand{\eaq}{\end{eqnarray}}
\newcommand{\beqs}{\begin{subequations}}
\newcommand{\eeqs}{\end{subequations}}
\newcommand{\bel}{\begin{align}}
\newcommand{\eal}{\end{align}}

\newcommand\eqs[2]{Eqs.~(\ref{#1}) and (\ref{#2})}

\newcommand\eqss[3]{Eqs.~(\ref{#1}), (\ref{#2}), and (\ref{#3})}

\newcommand{\ftn}{\footnotesize}

\newcommand{\ssz}{\scriptsize}
\newcommand{\TeV}{{\mbox{\rm TeV}}}

\newcommand{\EeV}{{\mbox{\rm EeV}}}
\newcommand{\PeV}{{\mbox{\rm PeV}}}
\newcommand{\ZeV}{{\mbox{\rm ZeV}}}
\newcommand{\YeV}{{\mbox{\rm YeV}}}

\newcommand{\etal}{{\it et al.\/}}

\def\to{\rightarrow}

\def\lf{\left(}
\def\rg{\right)}
\newcommand\vev[1]{\langle {#1} \rangle}
\newcommand\vevi[1]{\langle {#1} \rangle_{\rm I}}

\newcommand{\Nhi}{\ensuremath{N_{\rm I\star}}}
\newcommand{\ks}{\ensuremath{k_\star}}
\newcommand{\Gsm}{\ensuremath{\mathbb{G}_{\rm SM}}}

\newcommand{\Vhi}{\ensuremath{V_{\rm I}}}

\newcommand{\Hhi}{\ensuremath{H_{\rm I}}}
\newcommand{\Whi}{\ensuremath{W}}
\newcommand{\Khi}{\ensuremath{K_{\rm I}}}

\newcommand{\Vhio}{\ensuremath{V_{\rm I0}}}

\newcommand{\mP}{\ensuremath{m_{\rm P}}}

\newcommand{\Ggut}{\ensuremath{\mathbb{G}}}

\newcommand{\Gfl}{\ensuremath{\mathbb{G}_{\rm 51F}}}
\newcommand{\Glr}{\ensuremath{\mathbb{G}_{\rm LR}}}
\newcommand{\Gbl}{\ensuremath{\mathbb{G}_{B-L}}}

\newcommand{\ubl}{\ensuremath{U(1)_{B-L}}}

\newcommand{\aS}{\ensuremath{{\rm a}_S}}

\newcommand{\msn}{\ensuremath{m_{\rm I}}}

\newcommand{\Nr}{\ensuremath{{\sf N}_{\mathbb{G}}}}

\newcommand{\ns}{\ensuremath{n_{\rm s}}}
\newcommand{\as}{\ensuremath{\alpha_{\rm s}}}

\newcommand{\sni}{\ensuremath{\nu^c_i}}

\newcommand{\Dmx}{\ensuremath{\Delta_{\rm m\star}}}
\newcommand{\Dcx}{\ensuremath{\Delta_{\rm c\star}}}

\newcommand{\sg}{\ensuremath{\sigma}}
\newcommand{\sgx}{\ensuremath{\sigma_\star}}
\newcommand{\sgc}{\ensuremath{\sigma_{\rm c}}}
\newcommand{\sgm}{\ensuremath{\sigma_{\rm max}}}
\newcommand{\sgf}{\ensuremath{\sigma_{\rm f}}}
\newcommand{\sgn}{\ensuremath{\sigma_{\rm min}}}
\newcommand{\sgb}{\ensuremath{\bar\sigma_{\rm min}}}

\newcommand{\kp}{\ensuremath{\kappa}}
\newcommand{\aaS}{\ensuremath{|\aS|}}

\def\ssb{\leavevmode\hbox{$\diagup$\kern-12pt\ftn\scshape
susy}}

\newcommand{\fref}[1]{Fig.~\ref{#1}}

\newcommand{\Eref}[1]{Eq.~(\ref{#1})}
\newcommand{\Sref}[1]{Sec.~\ref{#1}}
\newcommand{\Fref}[1]{Fig.~\ref{#1}}
\newcommand{\Tref}[1]{Table~\ref{#1}}
\newcommand{\cref}[1]{Ref.~\cite{#1}}

\def\th{{\theta}}

\def\fhi{FHI~}
\def\bmp{BMP~}
\def\bmps{BMPs~}
\def\Ka{K\"{a}hler potential}
\def\Kaa{K\"{a}hler~}

\def\Kap{K\"{a}hler potential}

\def\bcp{{\sc\small Bicep2}/{\it Keck Array}}
\def\nano{{\sf\small NG15}}

\def\spt{{\sf\ftn P-ACT-SPT}}

\newcommand{\plk}{{\it Planck}}

\newcommand{\bdhh}{{\ensuremath{\normalsize I{\kern-2.9pt H}}}}

\newcommand{\phc}{\ensuremath{\Phi}}
\newcommand{\phcb}{\ensuremath{\bar\Phi}}

\def\al{{\alpha}}

\newcommand{\mcs}{\ensuremath{{G\mu}}}
\newcommand{\rcs}{\ensuremath{{r_{\rm cs}}}}
\newcommand{\ecs}{\ensuremath{{\epsilon_{\rm cs}}}}
\newcommand{\rms}{\ensuremath{{r_{\rm ms}}}}

\newcommand{\Ns}{\ensuremath{{N_{\rm I\star}}}}
\newcommand{\khi}{\ensuremath{K_{\rm I}}}

\newcommand{\csgr}{\ensuremath{C_{\mbox{\ssz\sc sugra}}}}
\newcommand{\cssb}{\ensuremath{C_{\mbox{\ssz\sc ssb}}}}
\newcommand{\crc}{\ensuremath{C_{\mbox{\ssz\sc rc}}}}

\def\bcp{B{\sc\small icep2}/{\it Keck Array}}

\def\Kap{K\"{a}hler potential}
\def\Km{K\"{a}hler manifold}
\def\Kaa{K\"{a}hler~}

\def\tmi{TMI}

\def\actc{{\sf\ftn P-ACT-LB-BK18}}
\def\spt{{\sf\ftn P-ACT-SPT}}

\renewcommand{\figurename}{\bf\small Figure}
\renewcommand{\tablename}{\bf\small Table}

\title{F-Term Hybrid Inflation with T-Model K\"ahler Geometry and Beyond}

\author{Waqas Ahmed,$^{\sf(1)}$ Constantinos Pallis,$^{\sf(2)}$ and Mansoor Ur Rehman$^{\sf(3)}$ \\
$^{\sf(1)}$ Center for Fundamental Physics and School of Artificial Intelligence, \\
Hubei Polytechnic University, Huangshi 435003, China; \\ E-mail: \email{waqasmit@hbpu.edu.cn}   \\
$^{\sf(2)}$ School of Technology,  Aristotle University of
Thessaloniki, \\ Thessaloniki, GR-54124 Greece;
\\ E-mail: \email{kpallis@auth.gr}\\
$^{\sf(3)}$ Department of Physics, Faculty of Science, Islamic
University of Madinah, \\ Madinah 42351, Saudi Arabia; \\ E-mail:
\email{m.rehman@iu.edu.sa}}

\abstract{We analyze F-term hybrid inflation (FHI) within various
grand unified theories in the presence of a K\"ahler potential for
the inflaton field which parameterizes the \Km s $SU(1,1)/U(1)$ or
$SU(2)/U(1)$.  We take into account supergravity, radiative, and
soft supersymmetry-breaking corrections to the tree-level
potential and find that viable FHI can be realized without extrema
along the inflationary trajectory for a broad region of the
parameter space. For selected superpotential parameters the
models' predictions are largely influenced by the curvature of the
internal space and the magnitude of the tadpole parameter which
are constrained so as to achieve compatibility with the current
ACT and SPT data. We also discuss the formation of cosmic strings
and their associated gravitational wave signals, potentially
detectable by current and upcoming experiments. }


\begin{document}





\section{Introduction}\label{intro}

Precision measurements of the \emph{Cosmic Microwave Background}
({\sf\ftn CMB}) have transformed inflation into a quantitatively
testable framework -- for a review see e.g. \cref{review}. The
Planck satellite has played a central role in this progress by
providing highly precise measurements of temperature and
polarization anisotropies over a wide range of angular scales.
Complementary ground-based experiments, such as the \emph{Atacama
Cosmology Telescope} ({\sf \small ACT}) and the \emph{South Pole
Telescope} ({\sf\small SPT}), have further improved constraints
through high-resolution observations of small-scale anisotropies.

While the results of these experiments are broadly consistent,
recent analyses reveal a mild tension in the preferred values of
the scalar spectral index, $\ns$. In particular, the combination
of {\it Planck} with \bcp\ yields \cite{plin,gws}
\begin{equation}
\label{nspl} n_{\rm s} \simeq 0.9652 \pm 0.0084
\quad\mbox{({\sf\ftn P-BK-LB} data at 95\%~\text{c.l.})},
\end{equation}
whereas the ACT \emph{data release 6} ({\sf\ftn DR6}), combined
with {\it Planck}, \bcp\, and DESI BAO measurements, prefers
\cite{act,actin}
\begin{equation}
n_{\rm s} = 0.9743 \pm 0.0068 \quad \Rightarrow \quad 0.967
\lesssim n_{\rm s} \lesssim 0.981\quad\mbox{(\actc\ data at
95\%~\text{c.l.})} \label{nsact}
\end{equation}
where the $\ns$ elevation above is mainly attributed to a tension
between CMB data and DESI BAO data \cite{tension}. On the other
hand, the SPT measurements, when combined with \plk\ and ACT data,
yield an intermediate value \cite{spt}
\begin{equation} n_{\rm
s} = 0.9684 \pm 0.006 \quad \Rightarrow \quad 0.962 \lesssim
n_{\rm s} \lesssim 0.974\quad\mbox{(\spt\ data at
95\%~\text{c.l.})}. \label{nsspt}
\end{equation}

{If the recent results above} -- mainly that in \Eref{nsact} --
will be confirmed by, e.g., future CMB data \cite{litebird, cmbs4,
so} they will place significant pressure on several well-motivated
\plk-consistent models such as \emph{T-Model inflation} ({\sf\ftn
TMI}) \cite{tmodel,alinde,disc}. TMI belongs to a class of
cosmological $\alpha$-attractors \cite{alinde}, which is
realized for inflaton values of the order of the reduced Planck
mass $\mP$ {close to a second-order pole emerging} in the inflaton
kinetic terms. This kinetic mixing may originate from a \Kap\ of
the form
\beq K_{\rm I} = -N\mP^2\ln\lf 1-|S|^2/N\mP^2\rg,  \label{khi}\eeq
where $N>0$ and $S$ is the gauge-singlet inflaton -- for \tmi\
with a gauge non-singlet inflaton see \cref{sor,tmdcs}. $\khi$
parameterizes the hyperbolic \Km\ $SU(1,1)/U(1)$ with constant
scalar (moduli space) curvature ${\sf R}_{\rm I}=-2/N$. This
mechanism leads to a robust and nearly universal prediction
$n_{\rm s}\simeq0.963$ for number of e-folding $\Ns\simeq55$
\cite{tmodel, ellis10}. Several variants
\cite{ellis10,gup,oxf,rha,act5,heur,waqas,laura,hz,actpole,phi}
have been proposed recently to reconcile TMI with the \actc\ data
in \Eref{nsact} -- for a review see \cref{actreview}.

In this work we wish to emphasize that beyond the strong regime
above the geometry of \tmi\ can be also applied in a weak regime
for $S$ values well below $\mP$. This regime is expected to
cooperate well with inflationary models which develop a plateau
due to the structure of their potential. {A prominent
representative of such models} is \emph{F-term hybrid inflation}
({\sf\ftn FHI}) \cite{hisusy} {which is undoubtedly a well-known
model} thanks to its observational flexibility and the ease of
embedding in particle models \cite{lect}. Most notably, FHI is
based on a renormalizable superpotential, uniquely {determined by
a gauge symmetry $\Ggut$ and a global $U(1)_R$ symmetry}; it does
not require fine-tuned superpotential parameters and
transplanckian inflaton values; it can be, moreover, naturally
followed by a \emph{Grand Unified Theory} ({\sf \ftn GUT}) phase
transition {which may lead to the production of cosmological
defects if predicted by the symmetry-breaking scheme} -- see e.g.
\cref{rachel}. Among them, \emph{Cosmic Strings} ({\sf\small CSs})
attract a fair amount of attention currently since they contribute
to the CMB and generate \cite{nano1} a stochastic background of
\emph{gravitational waves} ({\ftn\sf GWs}) in the nanohertz range,
probed by \emph{pulsar timing arrays} ({\ftn\sf PTAs}) such as
\emph{NANOGrav (15-yr)} \cite{nano}, \emph{EPTA} \cite{pta},
\emph{PPTA} \cite{ppta}, and \emph{CPTA}\cite{{cpta}}. The tension
of CSs is directly linked to the symmetry-breaking scale of
$\Ggut$, connecting ACT/SPT-preferred inflationary parameters with
the spectrum of the GWs observed by PTAs -- see, e.g.,
\cref{buch,Afzal,lrshafi,tmdcs,nasri,blfhi,actfhi1,antus}.

For a reliable approach to FHI, soft \emph{Supersymmetry}
({\sf\ftn SUSY})-breaking terms
\cite{sstad,sstad1,sstad2,sstad3,mfhi,kaihi,asfhi,asfhi1} and
\emph{Supergravity} ({\sf\ftn SUGRA}) corrections \cite{pana, gil,
mur, rlarge, rlarge1} have to be taken into account together with
the \emph{radiative corrections} ({\ftn \sf RCs}) employed in the
original version of the model \cite{hisusy}. {Both of the
aforementioned corrections are of crucial importance} in order to
reconcile the inflationary observables with present data
\cite{fhi3,fhi4,actfhi,actfhi1}. In particular, the SUGRA
corrections are obviously related to the adopted geometry of the
moduli space. In conventional treatments, canonical
\cite{mfhi,kaihi,asfhi,asfhi1,sstad,sstad1} or quasi-canonical
K\"ahler \cite{pana,mur,gil,rlarge,rlarge1} potentials  are
typically introduced to mitigate the $\eta$-problem of FHI -- for
a review see \cite{hinova}. However, such constructions often rely
on multiple small coefficients, which reduces the predictive power
of the framework and lacks a clear connection to known superstring
compactifications.

Inspired by the geometry of \tmi, we here adopt the \Kap\ in
\Eref{khi} and consider both signs of $N$. Note that, for $N<0$,
$\khi$ parameterizes the compact \Km\ $SU(2)/U(1)$ {with constant
scalar curvature as well}. This SUGRA setting is governed by just
one parameter, $N$ in \Eref{khi}, whereas FHI models based on
polynomial K\"ahler potentials typically require at least two
independent parameters. In this sense, the present construction is
more predictive. For $N>0$, $K_{\rm I}$ was first introduced in
\cref{hpana1,hpana2} in presence of a stabilized modulus to
alleviate the $\eta$ problem of FHI -- cf. \cref{kelar, smth}. It
was also shown that $N$ values close to the superconformal limit
can support FHI in \cref{nmfhi} in the context of flipped $SU(5)$
for specific values of the superpotential parameters. {It was also
employed, independently of inflation,} to accomplish
gravity-mediated SUSY breaking under the assumption of a mildly
violated $R$ symmetry \cite{susyr} or within no-scale SUGRA
\cite{nsde}.

We here consider $N$ as an input parameter which is restricted by
the inflationary requirements as in the case of \tmi\
\cite{tmodel,sor}. We find an ample available parameter space
adjusting (as functions of the superpotential parameters) $N$ in
conjunction with the tadpole parameter $\aS$ for $N>0$ or $N<0$.
In the major part of the emergent parameter space $|N|$ acquires
quite large values which do not pose, though, any observational --
{as in the case of \tmi} -- or theoretical difficulty. Indeed,
from the SUGRA perspective, $N$ is not fixed and may take
arbitrary values unless a particular ultraviolet completion is
specified -- see e.g. \cref{disc}. This situation may be compared
with that of non-minimal Higgs inflation
\cite{nm1,nm2,nm3,nm4,nm5,nmfhi}, where a very large coupling to
Ricci scalar curvature is typically required. In contrast to that
case, though, the $S$ values here are much lower than $\mP$ and so
$S$ turns out to be essentially canonically normalized. As a
consequence, $N$ does not appear in any numerator of the expansion
of the inflationary potential for low $S$ values and therefore,
{no issue with the} validity of the effective theory arises
\cite{un1,riotto}.



The salient features of our models are introduced in \Sref{fhi1}
and the resulting inflationary potential is derived in
\Sref{fhi2}. In \Sref{fhi3}, we show that the inflationary
observables can be successfully reconciled with a number of
constraints listed in \Sref{cons}. We also examine the formation
of CSs in one version of these models and discuss the associated
spectrum of GWs in \Sref{cssec}. Our conclusions are summarized in
\Sref{con}.

\section{Models' Setup}\label{fhi1}

The simplest version of FHI \cite{hisusy, hinova} can be
implemented by introducing three superfields $\bar{\Phi}$, $\Phi$
and $S$. {The first two are} left-handed chiral superfields
oppositely charged under a gauge group $\Ggut$ {while the latter
corresponds to the inflaton} and is a $\Ggut$-singlet left-handed
chiral superfield.  In this work we identify $\Ggut$ with three
possible gauge groups {with different representation dimensions
$\Nr$} to which $\bar{\Phi}$ and $\Phi$ belong -- see \Tref{tab1}.
Namely, we consider the following $\Ggut$'s
\beqs\bea \label{gbl} &\Gbl:= \Gsm\times U(1)_{B-L}\hspace*{3.8cm} &~~\mbox{with}~~\Nr=1,\\
\label{glr} &\Glr:= SU(3)_{\rm C}\times SU(2)_{\rm L} \times
SU(2)_{\rm R} \times U(1)_{B-L} &~~\mbox{with}~~\Nr=2,\\
\label{gfl} &\Gfl:= SU(5)\times U(1)_X \hspace*{4.05cm}
&~~\mbox{with}~~\Nr=10. \eea
Here \Gsm\ is the well-known gauge group of the \emph{Standard
Model} ({\sf\ftn SM})
\beq\label{gsm} \Gsm:= SU(3)_{\rm C}\times SU(2)_{\rm L} \times
U(1)_{Y},\eeq\eeqs
to which $\Ggut$ is broken via the \emph{vacuum expectation
values} ({\ftn\sf v.e.vs}) of $\Phi$ and $\bar\Phi$ at the end of
FHI. As regards the cosmological defects, CSs are produced only
for $\Ggut=\Gbl$ -- see \Sref{cssec}.

{Indeed, this scheme} can be realized if we adopt the
superpotential
\beq
W = \kp S\left(\bar \Phi\Phi-M^2\right),\label{Whi}
\eeq
where $\kp$ and $M$ are free parameters which may be constrained
by the inflationary requirements -- see \Sref{cons}. $W$ in
Eq.~(\ref{Whi}) is the most general renormalizable superpotential
consistent with a continuous $R$ symmetry \cite{hisusy} under
which
\begin{equation}
  \label{Rsym}
S\  \rightarrow\ e^{ir}\,S,~\bar \Phi\Phi\to\bar
\Phi\Phi~~\mbox{and}~~W\rightarrow\ e^{ir}\, W.
\end{equation}

{The SUSY potential, $V_{\mbox{\sc susy}}$, extracted from $W$ in
Eq.~(\ref{Whi}) includes contributions from F and D-terms. The
latter contribution vanishes along the direction
$|\bar\Phi|=|\Phi|$, whereas the former  (in the flat limit with
$\mP\to \infty$) reads
\beq \label{VF} V_{\rm F}=\mbox{$\sum_\al$}|{\rm F}_\al|^2=
\kappa^2\left((|\Phi|^2-M^2)^2+2|S|^2 |\Phi|^2\right), \eeq
where ${\rm F}_\al=\partial_{\phi_\al} W$ with
$\phi_\al=S,\phc,\phcb$ and the symbol $\partial_{X}$ {denotes the
partial derivative} \emph{with respect to} ({\small\sf w.r.t})
$X$.} Consequently, the SUSY minimum is
\beq\label{vevs} \vev{S} = 0 ~~ \mbox{and}~~|\vev{\Phi}| =
|\vev{\bar{\Phi}}| = M. \eeq
{At the vacuum} $\Ggut$ is spontaneously broken to $\Gsm$, and
SUSY is preserved up to soft SUSY-breaking terms -- see e.g.
\cref{tmdcs}. The inflaton system acquires mass
$\msn=\sqrt{2}\kp M$ \cite{lect}.

\renewcommand{\arraystretch}{1.3}
\begin{table}[t] \bec\begin{tabular}{|c|c|c|c|c|}\hline
{\sc Super-} &\multicolumn{3}{c|}{\sc Representations Under
$\Ggut$}&$R$\\\cline{2-4}
{\sc Fields}&\Gbl&\Glr&\Gfl&{\sc Charge}\\\hline\hline
\multicolumn{5}{|c|}{\sc Higgs Superfields}\\\hline
$\phc$&$({\bf 1, 1}, 0, 2)$&$({\bf 1, 1, 2}, 1)$&$({\bf 10}, 1)$&0\\
$\phcb$&$({\bf 1, 1}, 0, -2)$&$({\bf 1, 1, \bar 2}, -1)$&$({\bf
\overline{10}}, -1)$&0\\ \hline
$S$&$({\bf 1, 1}, 0, 0)$&$({\bf 1, 1, 1}, 0)$&$({\bf 1},0)$&$2$\\
\hline
\end{tabular}\eec
\caption{\sl\small Representations and $R$ charges of the
superfields involved in FHI for various $\Ggut$'s.}\label{tab1}
\end{table}
\renewcommand{\arraystretch}{1.}

{As mentioned in} \Sref{intro}, $\Whi$ in Eq.~(\ref{Whi}) may
cooperate with a variety of \Ka s for implementing FHI. We here
focus on the quadratic \Ka
\beq K =K_{\rm I}+ |\Phi|^2+|\bar\Phi|^2
\label{ki} \eeq
with $\Khi$ given in \Eref{khi}. {$K$  parameterizes the following
(curved) \Km:}
\beq (SU(1,1)/U(1))_S\times U(1)_{\phcb} \times
U(1)_\phc~~\mbox{for $N>0$ or}~~(SU(2)/U(1))_S\times U(1)_{\phcb}
\times U(1)_\phc~~\mbox{for $N<0$,} \eeq
where the subscripts indicate {the moduli parameterizing the
corresponding manifolds.}

\section{Inflationary Potential}\label{fhi2}

{It is well known \cite{hisusy} that FHI occurs for sufficiently
large values of $|S|$} along a F- and D- flat direction of the
F-term potential
\begin{equation} \label{v0}\bar\Phi={\Phi}=0,~~\mbox{where}~~ V_{\rm F}\lf{\Phi}=0\rg:= V_{\rm I0}=\kp^2
M^4~~\mbox{and}~~\Hhi=\sqrt{\Vhio/3\mP^2}\eeq
are the constant potential energy density and corresponding Hubble
parameter which drive FHI -- the subscript $0$ means that this is
the tree level value. The interplay of several corrections
determines {the slope} of the inflationary potential which can be
written as follows
\beq V_{\rm I}= V_{\rm I0}\lf 1+\csgr+\cssb+\crc\rg.\label{vhi}
\eeq
The individual contributions are specified below:

\subparagraph{\sf\ftn (a) SUGRA corrections}

Given that the D-term contributions into the SUGRA potential
vanish {due to the stabilization of} $\phc$ and $\phcb$ at the
origin, we focus on the F-term SUGRA scalar potential with just
one field $S$,
\beq V_{\mbox{\sc sugra}}=e^{K/\mP^2}\left(K^{SS^*}D_S \Whi
D^*_{S^*} \Whi^*-3{\vert
\Whi\vert^2/\mP^2}\right)\>\>\>\mbox{where}\>\>\>D_S = \partial_{S}
+  \partial_{S}K/\mP^2,\label{vsugra} \eeq
is the K\"{a}hler-covariant derivative. We also introduce the
inverse of the \Kaa metric
\beq K_{SS^*}=\partial_{S^*}\partial_S{K_{\rm
I}}=(1-\sg^2/2N\mP^2)^{-2}~~\mbox{with}~~S=\sg
e^{i\theta_S}/\sqrt{2} \label{kss}\eeq
using the relation $K^{SS^*}=K^{-1}_{SS^*}$. {Taking this result
into account} and plugging into \Eref{vsugra} the expression of
$\Khi$ in \Eref{khi}, we arrive at the following result
\beq V_{\mbox{\sc sugra}}= V_{\rm I0}\lf1 - \frac{\sg^2}{2
N\mP^2}\rg^{-N} \lf 1 - \frac{(2 + N)}{ N} \frac{\sg^2}{2\mP^2} +
\lf\frac{N-1}{N}\rg^2 \frac{\sg^4}{4\mP^4}\rg\,.\label{vsgre} \eeq
Expanding for $S\ll\mP$ the expression above, we can identify the
SUGRA corrections of \Eref{vhi} after subtracting the dominant
constant contribution of \Eref{v0}, i.e.,
\beq \csgr=-\frac{1}{N}\frac{\sg^2}{\mP^2}+\frac{2 + N(N-7) }{8
N^2}\frac{\sg^4}{\mP^4} +\frac{1 + N (2 N-9)}{24
N^2}\frac{\sg^6}{\mP^6}+\cdots \label{csgr}\eeq
{As can be seen from the} expansion above, the largest coefficient
of the various terms for $N\gg1$ is of order unity and so the
ultraviolet cut-off scale of our theory is $\mP$. In the same $N$
limit, we can reproduce the expansion obtained for canonical
$\Khi$ -- see e.g. \cref{hinova}.

\subparagraph{\sf\ftn  (b) Soft SUSY-breaking corrections.}
Possible coupling of the inflationary sector with the
SUSY-breaking sector -- cf. \cref{asfhi} -- generates corrections
to $\Vhi/\Vhio$ {arising from} soft SUSY-breaking terms. Namely,
these corrections can be written as
\beq V_{\mbox{\sc ssb}}=-\kp M^2 (\aS S+\aS ^* S^*)
~\Rightarrow~\vevi{V_{\mbox{\sc ssb}}}=\Vhio
\cssb~~\mbox{with}~~\cssb= -\sqrt{2}|\aS|\sg/\Vhio,\label{vssb}
\eeq
where the minus sign {arises from} the stabilization of
$\th=\theta_S+\theta_{\aS}$ -- see \Eref{kss}. Indeed, for
$\theta/\mP=0~({\sf mod}~2\pi)$ {the contribution} $-\aS S+ {\rm
c.c.}=-\sqrt{2}|\aS|\sg\cos(\theta/\mP)$ is minimized. Here we
further assume that $\theta_S$ remains constant during FHI so that
the simple one-field slow-roll approximation is valid. Possible
variation of $\theta_S$ is investigated in \cref{kaihi} {where it
was found that} acceptable solutions with $\theta\neq0$ {require
significant fine-tuning}. We also checked that a possible soft
SUSY-breaking mass term $m_S^2|S|^2$ with $m_S$ of the order of
$\aS$ is irrelevant for the values of the parameters considered in
our setup.

\subparagraph{\sf\ftn (c) Radiative corrections.}

These corrections originate \cite{hisusy} from a mass splitting in
the $\Phi-\bar{\Phi}$ supermultiplets due to SUSY breaking on the
inflationary valley. To compute them we work out the mass spectrum
of the fluctuations of the various fields about the inflationary
trough in \Eref{v0}. We obtain $2\Nr$ Weyl fermions and $2\Nr$
pairs of real scalars with mass squared respectively
\beq m_{\rm f}^2=\kp^2\sg^2/2
~~\mbox{and}~~m_{\pm}^2=\kp^2(\sg^2/2\pm M^2)~~\label{mscalar}\eeq
Inserting these masses into the well-known Coleman-Weinberg
formula, we find the correction
\beq V_{\mbox{\sc rc}}=\Vhio \crc~~\mbox{with}~~
\crc={\kp^2\Nr\over32\pi^2}\lf\sum_{i=\pm} m_{i}^4\ln{m_{i}^2\over
Q^2} -2m_{\rm f}^4\ln{m_{\rm f}^2 \over Q^2}\rg,\label{vrc}\eeq
where $Q$ is a renormalization scale. In the largest part of the
parametric space of the model $m_{-}^2$ develops a tachyonic
instability for
\beq \sg<\sgc=M/\sqrt{2}\label{sc}\eeq
terminating abruptly FHI. SUGRA corrections to the masses above
are at most of the order $M^4/\mP^2$ and can be safely ignored.

\section{Inflationary Requirements}\label{cons}

The parameters of FHI models {can be constrained by imposing a
number} of observational and theoretical constraints described in
Secs.~\ref{obs1} and \ref{obs2} below.

\newcommand{\dphp}{\ensuremath{\delta\Phi_+}}

\subsection{Observational Constraints}\label{obs1}

{Below we describe a number of observational constraints
considered in our investigation. In the relevant standard
formulas, we take into account that $\sg$ effectively coincides
with the canonically normalized inflaton since
$\vevi{K_{SS^*}}\simeq1$, thanks  to the facts that $\sg\ll\mP$ and
$N\gg1$. Namely:}

\subparagraph{\sf\ftn (a)} The number of e-foldings that the pivot
scale $\ks=0.05/{\rm Mpc}$ suffered during FHI have to be enough
to resolve {the shortcomings of} the Standard Big Bang, i.e.,
\cite{plin}:
\begin{equation}  \label{nhi}
 N_{\rm I*}=\frac{1}{m^2_{\rm P}}\;
\int_{\sgf}^{\sgx}\, d\sigma\: \frac{V_{\rm I}}{V'_{\rm
I}}\simeq\frac{1}{\mP^2}\int_{\sgf}^{\sgx}\,
\frac{d\sigma}{\crc'+\cssb'+\csgr'} \simeq19.4+{2\over
3}\ln{V^{1/4}_{\rm I0}\over{1~{\rm GeV}}}+ {1\over3}\ln {T_{\rm
rh}\over{1~{\rm GeV}}},
\end{equation}
where prime denotes derivation w.r.t $\sg$, $\sgx$ is the value of
$\sigma$ when $\ks$ {exits the horizon} of FHI and $\sigma_{\rm
f}\simeq\sgc$ signals the termination of FHI due to the tachyonic
instability encountered in \Eref{vrc}. Despite this well-known
fact, we also check the validity of slow-roll condition which
reads
\beq \label{slow} {\sf\ftn
max}\{\epsilon(\sgf),|\eta(\sgf)|\}=1,\eeq
where the relevant slow-roll parameters can be found from the
relations
\beq \label{sr} \epsilon\simeq{\mP^2\over2}\left(\frac{V'_{\rm
I}}{V_{\rm I}}\right)^2\simeq \frac{\mP^2}{2}\lf
 \crc'+  \cssb'+  \csgr'\rg^2~~\mbox{and}~~\eta\simeq m^2_{\rm P}~\frac{V''_{\rm
I}}{V_{\rm I}}\simeq\mP^2 \lf \crc''+\csgr''\rg. \eeq
{In the last part of \Eref{nhi}, we assume that \fhi\ is followed
in turn by an oscillatory phase with equation-of-state parameter
$w_{\rm rh}=0$, radiation and matter domination. Note that after
FHI we obtain two oscillatory systems composed by the complex
scalar fields $S$ and $\dphp=(\delta\bar \Phi+\delta
\Phi)/\sqrt{2}$, where $\delta\bar \Phi=\bar \Phi-M$ and
$\delta\Phi=\Phi-M$. Given that $w_{\varphi}=(p-2)/(p+2)$ for a
power-law potential $\varphi^p$ \cite{turner}, we expect that
$w_S\simeq0$ and $w_{\Phi}\simeq1/3$ since $V_{\rm F}$ in
\Eref{VF} displays a quadratic [quartic] dependence on $S$
[$\Phi$] with $\Phi$ [$S$] close to its value in \Eref{vevs}. As
shown in the related setting of \cref{smth}, the energy density of
$\dphp$ is overshadowed by that of $S$ which, thereby, dominates
the reheating phase yielding $w_{\rm rh}=w_S\simeq0$. Lastly}, in
\Eref{nhi} {we adopt for the reheating temperature} $T_{\rm rh}$ a
value close to $(0.01-1)~\EeV$. {These values are} compatible with
non-thermal leptogenesis \cite{lept,zhang} {which can be naturally
accommodated within FHI} -- cf. \cref{mfhi,nmfhi,sstad3}. In all
cases, we obtain $\Nhi\simeq50$.

\subparagraph{\sf\ftn (b)} The amplitude $A_{\rm s}$ of the power
spectrum of the curvature perturbation generated by $\sigma$
during FHI must be appropriately normalized \cite{actin}, i.e.,
\begin{equation} \label{prob}
A_{\rm s}= \frac{1}{12\, \pi^2 m^6_{\rm P}}\left.\frac{V_{\rm
I}^{3}}{|V'_{\rm I}|^2}\right|_{\sg=\sgx}\simeq\frac{1}{12\, \pi^2
m^6_{\rm
P}}\left.\frac{\Vhio}{(\crc'+\cssb'+\csgr')^2}\right|_{\sg=\sgx}\simeq2.1326\times
10^{-9}.
\end{equation}

\subparagraph{\sf\ftn (c)} The remaining observables -- the scalar
spectral index $\ns$, its running $\as$, and the scalar-to-tensor
ratio $r$ -- are calculated by the standard slow-roll formulas
\begin{equation} \label{ns}
\ns = 1 - 6 \epsilon_\star + 2 \eta_\star, ~~\as =
\frac{2}{3}\left(4\eta_\star^2 - (\ns-1)^2\right) - 2
\xi_\star~~\mbox{and}~~ r = 16 \epsilon_\star,
\end{equation}
where $\xi_\star \simeq \mP^4V'_{\rm I} V'''_{\rm I} / V_{\rm
I}^2$, and  all quantities with the subscript $\star$ are
evaluated at $\sigma = \sgx$. These observables must be consistent
with current CMB measurements. {Using the data from}
\eqs{nsact}{nsspt} and considering the $\Lambda$CDM+$r$ model we
end up with the following allowed range
\begin{equation}
0.962 \lesssim \ns \lesssim 0.981 ~~\mbox{and}~~r \lesssim 0.038,
~~\text{at 95\% c.l.} \label{nstot}
\end{equation}
{with a negligible running of $\ns$, namely $\as$.}

\subsection{Theoretical Considerations}\label{obs2}

From a more theoretical point of view, {our model can be further
refined} using the following criteria:

\subparagraph{\sf\ftn (a) Boundedness of $\Vhi$.} Requiring that
$\Vhi$ is bounded from below, {we leave open the possibility} that
FHI may occur under generic initial conditions set at
$\sigma\simeq\mP$. Since the large $\sg$ behavior of $\Vhi$ is
predominantly controlled by the positive (for large $N$ values)
higher order terms in \Eref{csgr}, the boundedness of $\Vhi$ {is
naturally ensured} in our scheme.

\subparagraph{\sf\ftn (b) Convergence of $\Vhi$.} The {expansion}
of $\Vhi$ in \Eref{csgr} is expected to converge at least for
$\sg\sim\sgx$. This {property} is automatically ensured in our
set-up thanks to the specific form of $\khi$ in \Eref{khi} which
exhibits a convergent expansion for $\sg\ll\mP$ and $|N|>1$. This
{advantage} of our proposal has to be contrasted with the
polynomial $\khi$ \cite{rlarge,rlarge1,hinova} which include
multiple coefficients of order unity in $\khi$ and so the
convergence of the expansion of $V_{\mbox{\sc sugra}}$
\Eref{vsugra} has to be checked in each case separately.

\subparagraph{\sf\ftn (c) Monotonicity of $\Vhi$.} Depending on
the values of $k$, $N$ and $\aS$  in \Eref{vhi}, $\Vhi$ is a
monotonic function of $\sigma$ or develops a local minimum and
maximum. The latter case leads to the {possibility} in which the
system gets trapped near the minimum of the inflationary potential
and, consequently, no FHI takes place. It is, therefore, crucial
to check if we can avoid the minimum-maximum structure of $V_{\rm
I}$. In such a case the system can start its slow rolling from any
point on the inflationary path without the danger of getting
trapped. This can be achieved if we require that $V_{\rm I}$ is a
monotonically increasing function of $\sigma$, i.e. $V'_{\rm I}>0$
for any $\sigma$ or, equivalently,
\beq V'_{\rm I}(\sgb)>0~~\mbox{with}~~ V''_{\rm
I}(\sgb)=0~~\mbox{and}~~V'''_{\rm I}(\sgb)>0,\label{cond}\eeq
where $\sgb$ {denotes} the value of $\sigma$ at which the minimum
of $V'_{\rm I}$ lies.

\subparagraph{\sf\ftn (d) Tuning of the initial conditions.} When
hilltop FHI occurs with $\sigma$ rolling from the region of the
maximum down to smaller values, a mild tuning of the initial
conditions is required \cite{hinova} in order to obtain acceptable
$n_{\rm s}$'s. In particular, the {smaller the desired value of
$n_{\rm s}$}, the closer we must set $\sgx$ to $\sgm$, where
$\sgm$  is the value of $\sg$ at which the maximum of $V_{\rm I}$
lies. On the other hand, in cases where $\aS$ plays a crucial
role in the inflationary dynamics -- especially for $N<0$ --
$\sgx$ must lie close to $\sgc$, giving rise to another source of
possible tuning. To quantify the amount of tuning involved, we
define \cite{mfhi,asfhi} the quantities:
\beq \Dmx=\left(\sgm -\sgx\right)/\sgm~~\mbox{and}~~\Dcx=\left(\sgx
-\sgc\right)/\sgx.\label{dms}\eeq
{The natural realization of FHI improves as $\Dmx$ and
$\Dcx$ increase.}

\subparagraph{} Let us, finally, note that in (standard) FHI --
cf. \cref{smth} -- the $\Ggut$-breaking scale $M$ is constrained
by \Eref{prob} and in most cases {it is typically slightly lower}
than the value dictated by the unification of the gauge coupling
constants within \emph{Minimal SUSY SM} {{\sf\ftn  MSSM}), i.e.,
\cite{smth,hinova}
\beq \label{Mgut} {g M}\simeq2 \times
10~\YeV\>\Rightarrow\>M\simeq28.6~\YeV~~\mbox{with}~~g\simeq0.7,\eeq
being the value of the unified gauge coupling constant. In our
setup we do not fulfill \Eref{Mgut}. This is totally acceptable
for $\Ggut=\Gbl$ since the gauge boson associated with the $\ubl$
breaking is neutral under $\Gsm$ and so it does not contribute to
the relevant renormalization group running -- this case can be
further constrained, though, as discussed in \Sref{cssec}. On the
other hand, if $\Ggut=\Glr$ or $\Gfl$ we may invoke threshold
corrections or additional matter supermultiples to restore the
gauge coupling unification -- for $\Ggut=\Glr$ see
\cref{tmdcs,lrshafi} and for $\Ggut=\Gfl$ see \cref{ggfl}.

\section{Inflation Analysis}\label{fhi3}

{We first analyze analytically}, in \Sref{fhi3a}, and then
numerically, in \Sref{fhi3b}, the inflationary dynamics of our
model.

\subsection{Analytic Results}\label{fhi3a}

For $\sgx\gg\sgc$ the contributions $\crc$ and $\csgr$ to $\Vhi$
in \Eref{vhi} can be approximated as
\beq
\crc\simeq\frac{\kp^2\Nr}{16\pi^2}\lf\frac{\kp\sg^2}{2Q^2}+\frac32\rg~~\mbox{and}~~
\csgr\simeq-\frac{1}{N}\frac{\sg^2}{\mP^2}+\frac18\frac{\sg^4}{\mP^4}.\label{csap}
\eeq
The relevant derivatives w.r.t $\sg$ which are involved in the
calculation of the various inflationary observables of \Sref{obs1}
read
\beqs\bel\label{cdev3}
\csgr'\simeq-\frac{2}{N}\frac{\sg}{\mP^2}+\frac12\frac{\sg^3}{\mP^4}~~&\mbox{and}~~
\csgr''\simeq-\frac{2}{N\mP^2}+\frac32\frac{\sg^2}{\mP^4},\\
\label{cdev2}
\crc'\simeq\frac{\kp^2\Nr}{8\pi^2}\frac{1}{\sg}~~&\mbox{and}~~
\crc''\simeq-\frac{\kp^2\Nr}{8\pi^2}\frac{1}{\sg^2},\\
\label{cdev1} \cssb'\simeq-\sqrt{2}\aS/\sqrt{\Vhio}~~&\mbox{and}~~
\cssb''=0.
\end{align}
\eeqs
From the expressions above we can easily verify that in the limit
of minimal SUGRA -- where $N$ tends to infinity and so the
quadratic term in $\csgr$ tends to zero -- $\crc''$ is the unique
contribution to $\eta$ in \Eref{sr} {thereby driving} $\ns$ to an
unacceptably large value \cite{hisusy, actfhi} keeping $\Vhi$
monotonic. {A careful choice of} $\aaS$ as a function of $\kp$
\cite{mfhi,kaihi,actfhi1} renders $\Vhi$ non-monotonic and reduces
$\ns$ to an acceptable level. {Taking advantage of the}
expressions above {we can easily infer that} $\Vhi'''>0$ for any
$\sg>0$ and so the monotonicity criterion in Eq.~(\ref{cond}) --
which is closely related to the $\ns$ values -- reduces to the
following condition
\beq \Vhi'(\sgb)>0~~\Rightarrow~~ \frac{\kp^3M^2\Nr}{\pi^2\sgb} -
4 \kp M^2\frac{4  \mP^2 \sgb - N \sgb^3}{N\mP^4} \geq 8 \sqrt{2}
\aS, \label{vmon}\eeq
where $\sgb$ can be derived by solving the equation in
\Eref{cond}. This task can be simplified, if we analyze separately
the situation emerging for the two signs of $N$. Namely:

\subsubsection{$N>0$ Case.} In this case, the reduction of $\ns$ at the
level of \Eref{nstot} can be {achieved} by increasing the negative
coefficients of $\sg^2$ in $\csgr$ and/or of $\sg$ in $\cssb$. In
both cases the monotonicity of $\Vhi$ in \Eref{vhi} can be
investigated in the same manner by applying \Eref{cond}. Namely,
{the value of $\sgb$ can be approximately determined} by solving
the equation
\beq \Vhi''(\sgb)=0~~\Rightarrow~~\sgb\simeq\frac{\mP}{\sqrt{6\pi}}\sqrt{\sqrt{3\kp^2\Nr+\frac{16\pi^2}{N^2}}+\frac{4\pi}{N}}
\sim  \frac{2}{\sqrt{3
N}}\mP,\label{sbp}
\eeq
where we use the rightmost expression in \Eref{cdev3} and neglect
the contribution from $\crc''$ in \Eref{cdev2} for the derivation
of the last result above. Note that the total result is
$\aS$-independent due to the rightmost result in \Eref{cdev1}. {If
$\Vhi$ is not monotonic}, it reaches a maximum at $\sg=\sgm$ which
can be {accurately} approximated as follows
\beq\label{smp} \Vhi'(\sgm)=0~~\Rightarrow~~\sgm=\frac{N\mP^2}{4}
\left( -\frac{\sqrt{2}|\aS|}{\kappa M^2} +
\sqrt{\frac{2|\aS|^{2}}{\kappa^{2} M^{4}} +
\frac{\kappa^{2}\Nr}{N\pi^{2}\mP^{2}}}\right),
\end{equation}
where the term of $\csgr$ in \Eref{csgr} {proportional to} $\sg^4$
has been ignored for simplicity.

The investigation of the remaining inflationary features {is
simplified if we distinguish} two subcases depending on the
{magnitude} of $\cssb$. Namely, we consider the cases:

\subparagraph{\sf{\ftn (a)} $N>0$ {\ftn Without Sizable} $\cssb$.}
In this case $\cssb$ can be totally neglected in \Eref{vhi} and so
$\Vhi$ can be simplified appreciably. When \Eref{vmon} is
violated, $\Vhi$ develops a maximum and minimum {for the following
values of $\sg$, respectively}
\beq \label{smn0} \sgm\simeq
\frac{\kappa\sqrt{N\Nr}}{4\pi}\mP~~\mbox{and}~~\sigma_{\rm
min}\simeq \frac{2}{\sqrt{N}}\mP.\eeq
Note that $\sgm$ is dominated by $\crc'$ in \Eref{cdev2} and the
quadratic term of $\csgr$ in \Eref{csgr} with the derived result
being consistent with \Eref{smp} in the limit $\aaS=0$. On the
other hand, $\sgn$  {is approximately determined by the vanishing}
of $\csgr'$ since $\csgr$ in \Eref{csgr} dominates the large-$\sg$
behavior of $\Vhi$.

Performing the integration in \Eref{nhi} for $\sgx\gg\sgf$ and
expanding the resulting expression in powers of $\sgx$ we can
reliably estimate $\Ns$ with the result
\beq\label{nhian}\Ns\simeq\frac{4\pi^2 \sgx^2} {\kp^4 \mP^4
N\Nr^2} \lf\kp^2 \mP^2 N \Nr + 8\pi^2
\sgx^2\rg~~\Rightarrow~~\sgx\simeq\frac{\kp \mP}{4\pi} \sqrt{D_N
\Nr},\eeq
where $D_N=\sqrt{N(8\Ns + N)}-N$. If we substitute {the expression
for} $\sgx$ into \Eref{ns} we can obtain {sufficiently accurate}
expressions for the inflationary observables. In particular, we
obtain
\beq\label{nsan}\ns\simeq1-\frac{4}{N}-\frac{4}{D_N}+6\kp^2\frac{\Nr
D_N}{32\pi^2},\eeq
whereas the expressions for $\as$ and $r$ are more complicated
{resulting in sufficiently low values} which are comfortably
consistent with \Eref{nstot}.

\subparagraph{\sf{\ftn (b)} $N>0$ {\ftn With Sizable} $\cssb$.} In
this case, $\cssb$ in \Eref{vssb} {contributes significantly} to
$\Vhi$ -- see \Eref{vhi} -- and cannot be ignored. This effect,
though, {is restricted to specific} $\kp$ (or $M$) values. Its
presence reinforces the negative contribution of the first term in
$\csgr'$ -- see \Eref{cdev3} -- into $\Vhi'$. Therefore, $\aaS$
influences $\epsilon$ and consequently $\Nhi$ in \eqs{sr}{nhi}
respectively. Its presence in the relevant formulas complicates
the analytical computation and so, {it is not possible} to find
specific $\sgx=\sgx(\Nhi)$ and $\ns=\ns(\Nhi)$ relations -- as in
\eqs{nhian}{nsan} respectively. Consequently, we are obliged
to trust the results of our numerical calculation presented in
\Sref{fhi3b} below. As shown there, the ramifications due to
$\cssb$ for selected $\aaS$ are confined to isolated values of
$\kp$.

\subsubsection{$N<0$ Case.} In this case, the presence of
$\cssb$ in \Eref{vhi} is crucial for the viability of our
inflationary scenario since {this is the only negative
contribution} into $\Vhi$ that may assist in reducing $\ns$ to the
{acceptable range of} \Eref{nstot}. The determination of $\sgb$
now requires consideration of all terms in \Eref{vhi} with the
approximations in \Eref{csap}, i.e.,
\beq \label{sbn}
\Vhi''(\sgb)=0~~\Rightarrow~~\bar\sigma_{\rm
min}\simeq\frac{\mP}{\sqrt{6\pi}}\sqrt{\sqrt{3\kp^2\Nr+\frac{16\pi^2}{|N|^2}}-\frac{4\pi}{|N|}}.\eeq
As in \Eref{sbp}, we note that $\sgb$ is still independent of $\aaS$. By setting $N=-|N|$ in \Eref{vmon} we can obtain the
condition for which $\Vhi$ remains monotonic. In the opposite
case, the extrema of $\Vhi$ can be well approximated if we ignore
the quartic term in $\csgr$. Namely, we obtain
\beq \label{smn}
\Vhi'(\sg)=0~~\Rightarrow~~\sigma_{_{\genfrac{}{}{0pt}{}{\rm
max}{\rm min}}}=\frac{|N| m_{\rm P}^{2}}{4}
\left(\frac{\sqrt{2}|\aS|}{\kappa M^{2}} \mp \sqrt{
\frac{2|\aS|^{2}}{\kappa^{2} M^{4}} -\frac{\kappa^{2}
\Nr}{|N|\pi^{2}\mP^{2}} } \right).
\eeq
For $\sg>\sgn$, the  boundedness of $\Vhi$ is assured as
explained in \Sref{obs2}.

The numerical analysis in \Sref{fhi3b} for $N<0$ reveals
that, in most of the allowed parameter space, the successful
realization of FHI requires $\sgx$ to lie close to $\sgc$
\cite{mfhi}. As a consequence, the expansion of $\crc$ in
\Eref{csap} ceases to be sufficiently accurate. The appropriate behavior of $\Vhi$, for given $\Nr$ and
$N$, can be achieved thanks to the similar magnitudes and opposite
signs of the terms $\crc'+\csgr'$ and $\cssb'$ in
\eqss{cdev3}{cdev2}{cdev1} which {can be achieved through a
careful choice} of $\kp$ and $\aS$. This arrangement accommodates
large enough $\Nhi$ -- see \Eref{nhi}. On the other hand, the
establishment of the inequality $|\crc''|>\csgr''$ {helps obtain}
$\eta<0$ and so low enough $\ns$ -- see \eqs{sr}{ns}.


\begin{figure}[!t]
    \centering
    \begin{subfigure}[b]{0.45\textwidth}
        \centering
        \includegraphics[width=\textwidth]{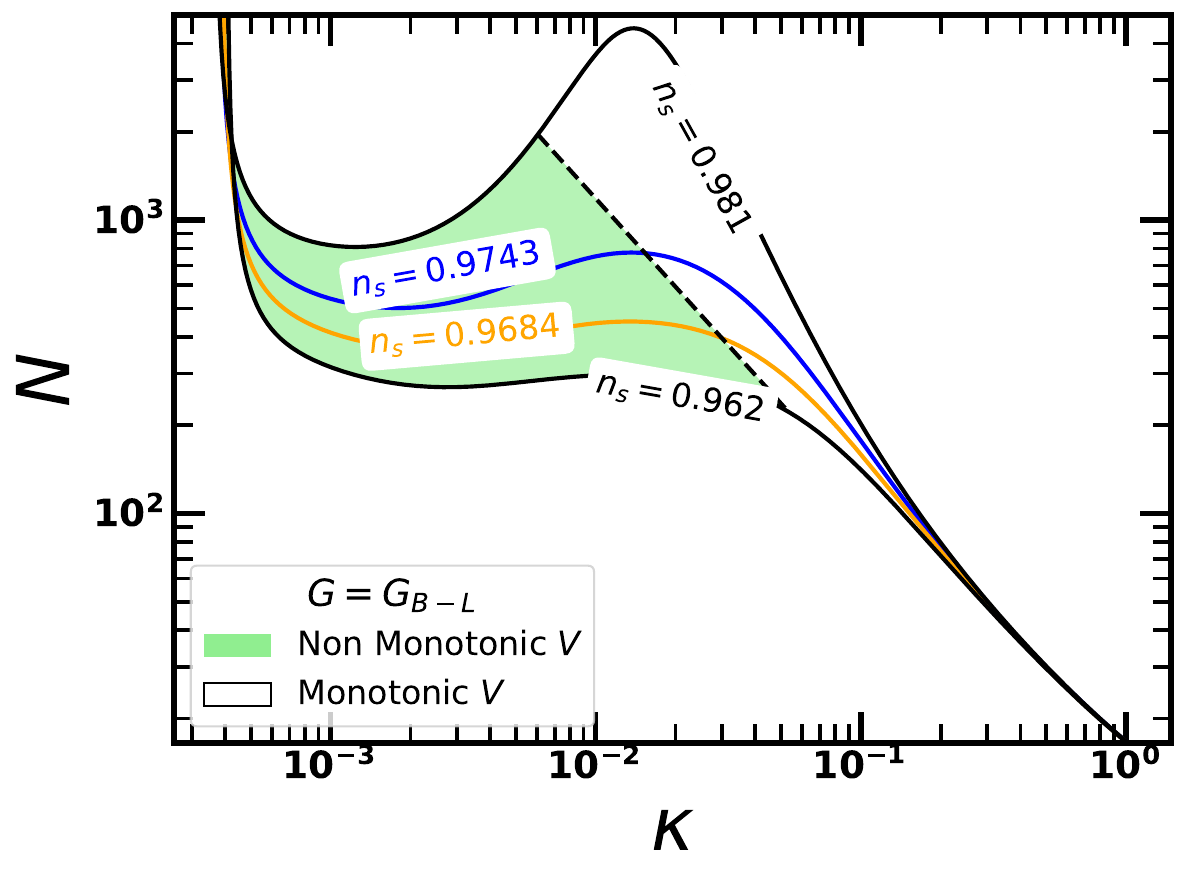}
    \end{subfigure}
    \hspace{0.0091\textwidth} 
    \begin{subfigure}[b]{0.45\textwidth}
        \centering
        \includegraphics[width=\textwidth]{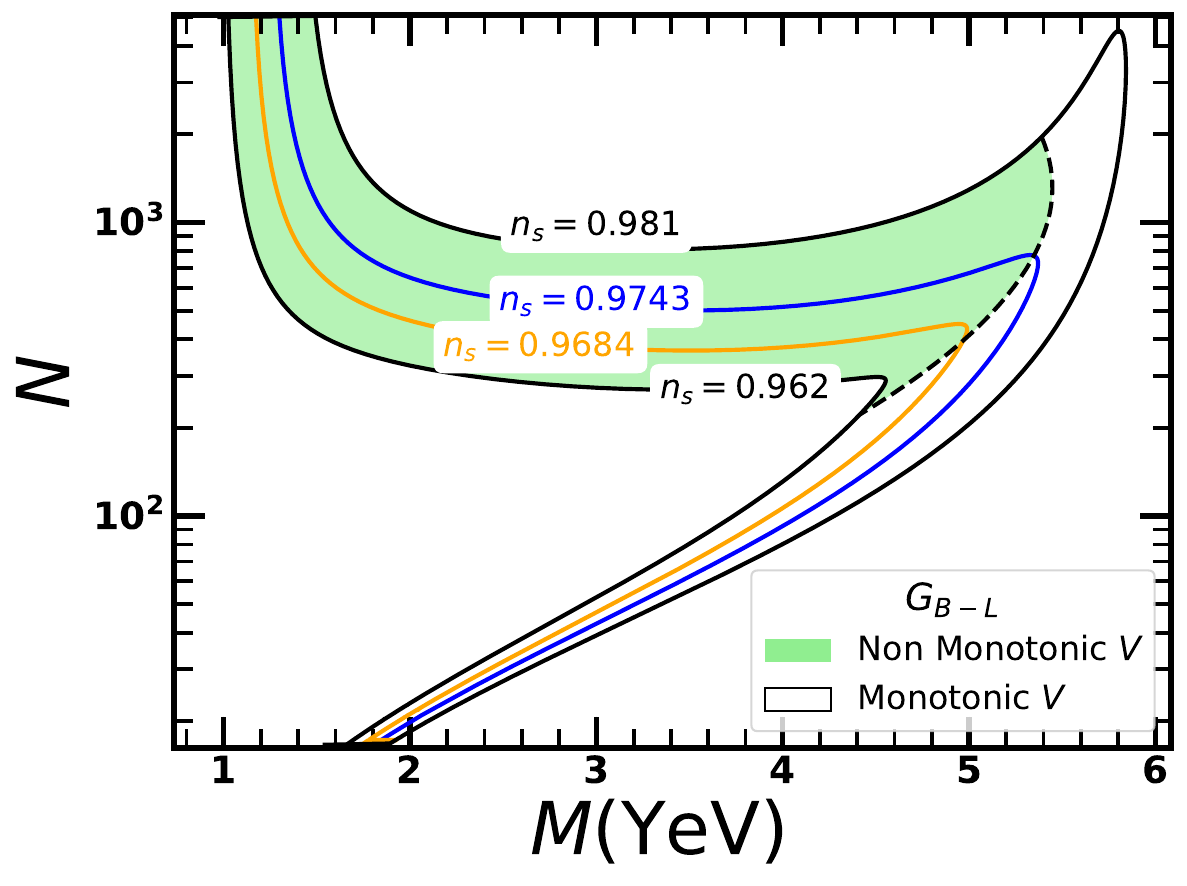}
    \end{subfigure}
    \begin{subfigure}[b]{0.45\textwidth}
        \centering
        \includegraphics[width=\textwidth]{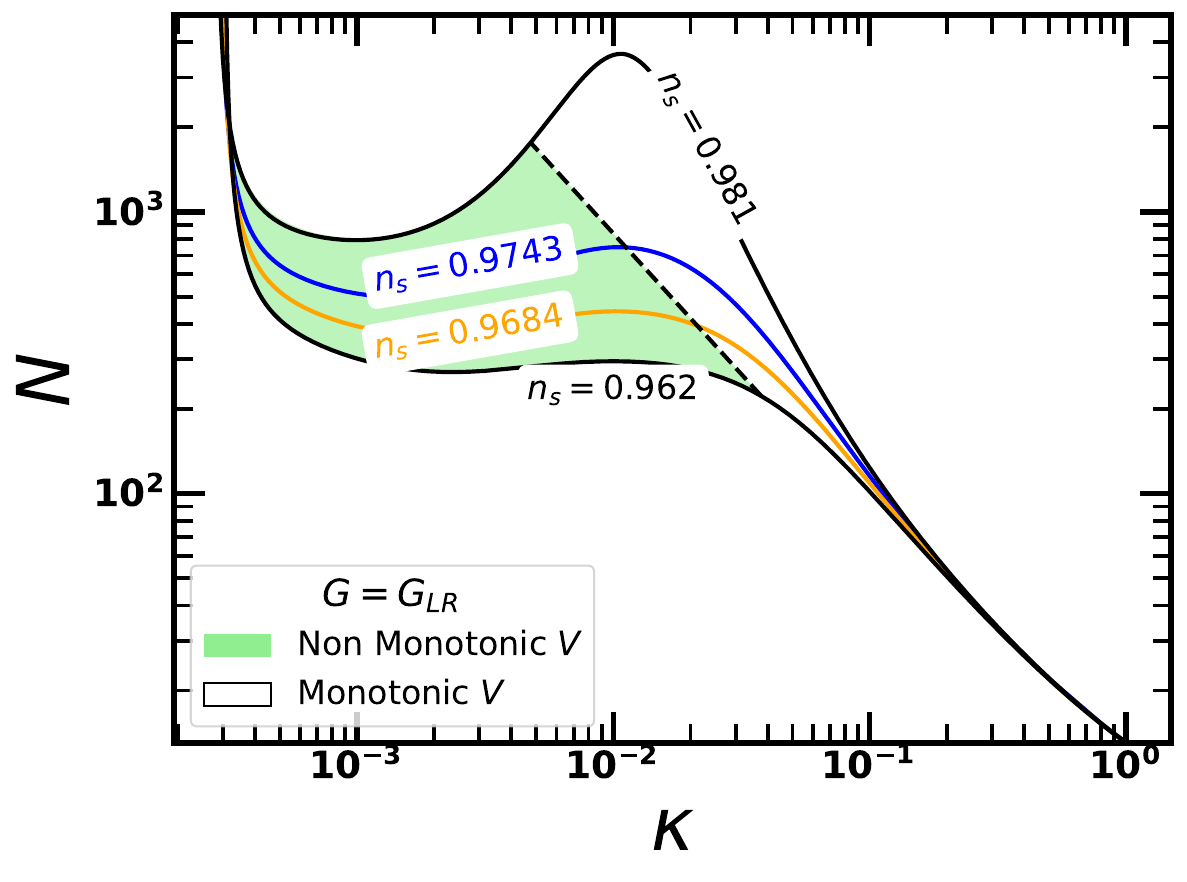}
    \end{subfigure}
 \hspace{0.0091\textwidth} 
    \begin{subfigure}[b]{0.45\textwidth}
        \centering
        \includegraphics[width=\textwidth]{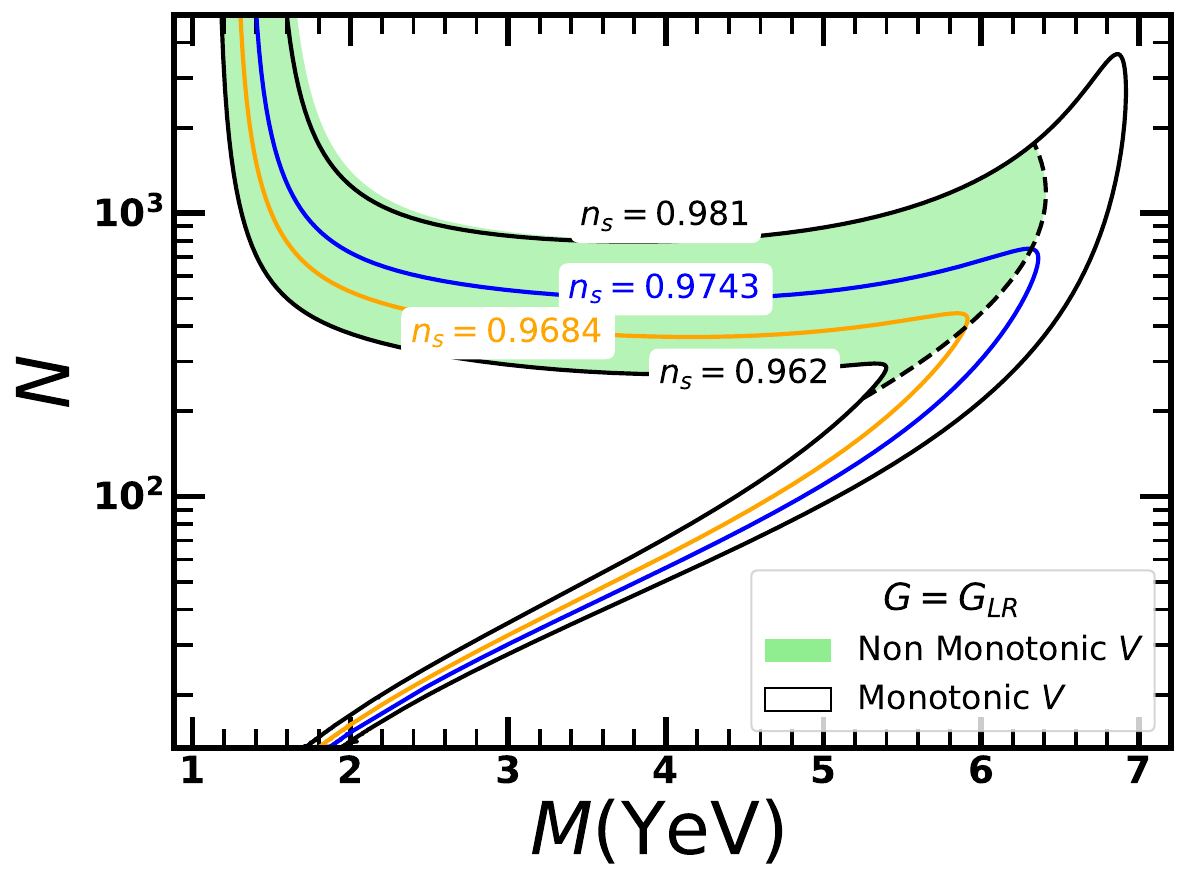}
    \end{subfigure}
\begin{subfigure}[b]{0.45\textwidth}
        \centering
        \includegraphics[width=\textwidth]{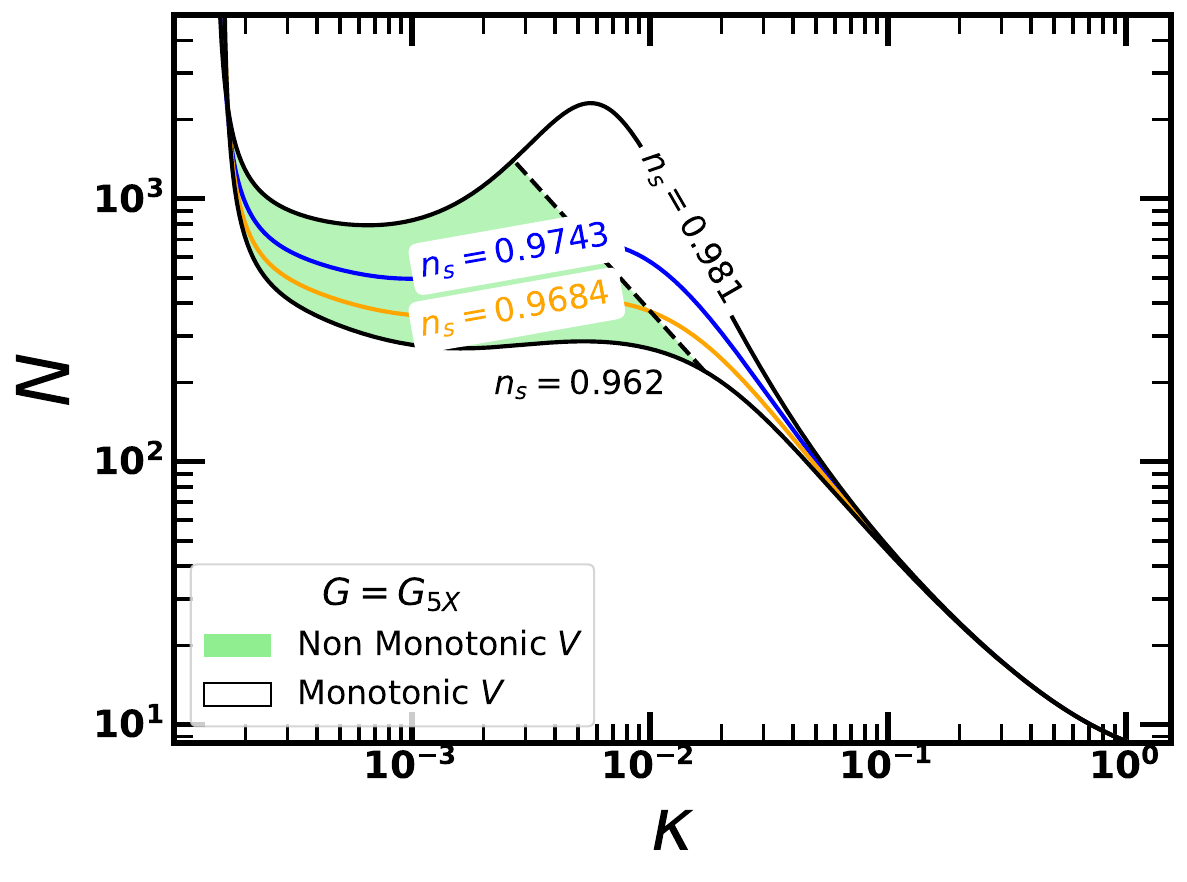}
    \end{subfigure}
         \begin{subfigure}[b]{0.45\textwidth}
        \centering
        \includegraphics[width=\textwidth]{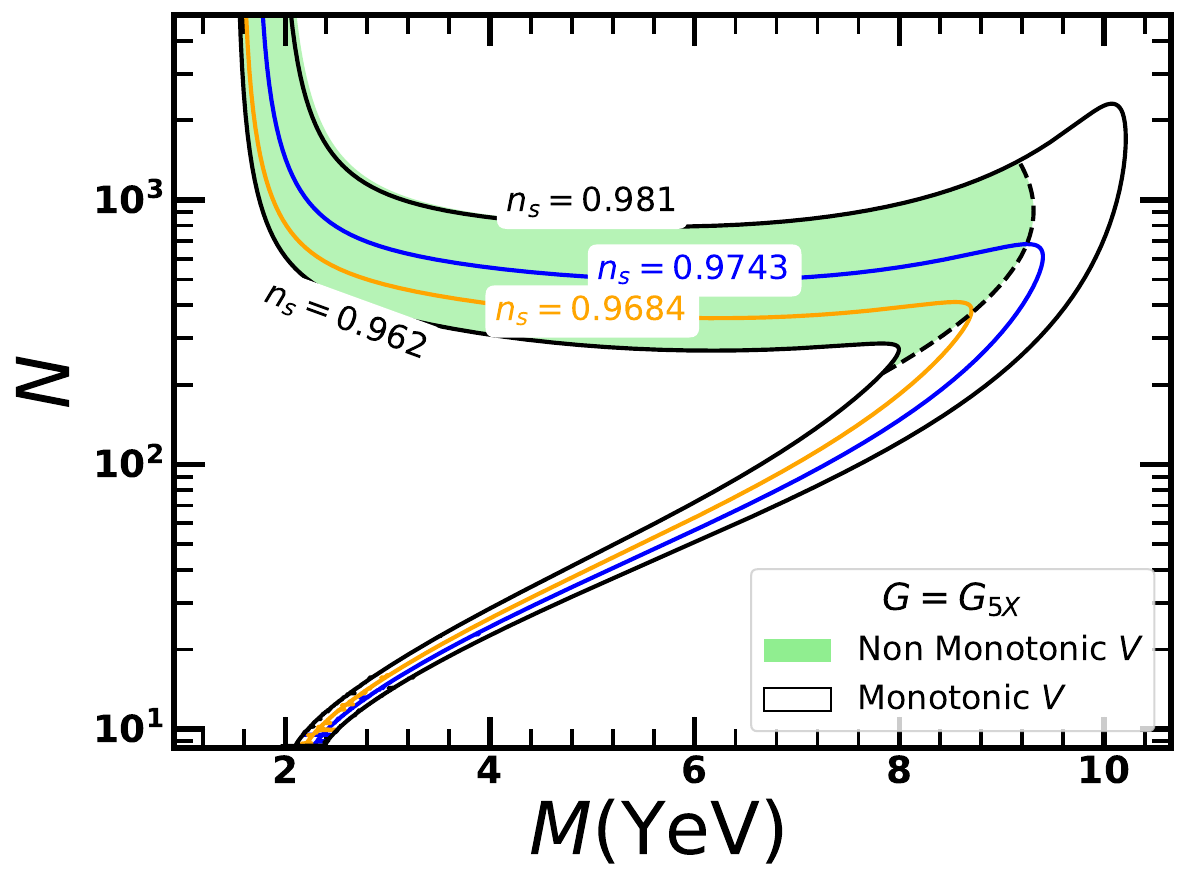}
    \end{subfigure}
\caption[]{\sl\small Allowed regions in $\kp-N$ (left panels)  and
$M-N$ (right panels) as determined by \eqss{nhi}{prob}{nstot} for
$|\aS| = 0.5~\text{TeV}$ and various $\Ggut$'s shown in the
legends. The light green region indicates the allowed parameter
space where $\Vhi$ is not monotonic whereas it remains monotonic
in the white patch. The dark blue [orange] curve yields the $\ns$
central value in \Eref{nsact} [\Eref{nsspt}]. } \label{fig2}
\end{figure}


\subsection{Numerical Results}\label{fhi3b}

We here present a detailed numerical analysis of the parameter
space of our model paying special attention {to identifying
regions} which ensure the monotonicity of $\Vhi$. Our strategy
{consists of restricting} $\sgx$ and $N$ as functions of our free
parameters $\kappa$ (or $M$) for given $\aS$ and $\Nr$, enforcing
Eqs.~(\ref{nhi}), (\ref{prob}) and (\ref{nstot}). For presentation
purposes we divide the analysis into the various cases introduced
in \Sref{fhi3a}. In particular:

\subparagraph{\sf{\ftn (a)} $N>0$ {\ftn Without} $\aaS$ {\ftn
Dependence}.} Our results are displayed in \Fref{fig2} where we
depict allowed regions in $\kp-N$ (left panels) and $M-N$ (right
panels) planes as determined by the aforementioned constraints for
$|\aS| = 0.5~\TeV$ and various $\Ggut$'s shown in the legends. The
lower [upper] boundary curves of the allowed regions come from the
lower [upper] bound on $\ns$ in \Eref{nstot}. {The dark blue
[orange] lines included within the allowed regions} correspond to
the central value of $\ns$ in \Eref{nsact} [\Eref{nsspt}].
Moreover, the light green region indicates the portion of the
allowed parameter space where $\Vhi$ is not monotonic while the
unshaded regions correspond to a monotonic $\Vhi$. The separation
of the two regions is indicated by the black dashed line, along
which the inequality in \Eref{vmon} is saturated.

\renewcommand{\arraystretch}{1.2}
\begin{table}[!t]
\begin{center}
\begin{tabular}{|l||cc|cc|cc|}
\hline
{\sf BMP}& {\sf A}&{\sf B}&{\sf C}&{\sf D}&{\sf E}&{\sf F}\\
\hline \hline
$\kp$ & $0.01$&{$0.01$}&$0.01$&$0.03$&$0.001$&$0.001$\\
$N$&$600$&$900$&$850$&$850$&$-500$&$-200$\\
$\aS/\TeV$&$1$&$1$&$5\times10^3$&$5\times10^3$&$45$&$54$\\\hline
$M/\YeV$& $6$&$6.3$&$5.5$&$6.5$&$2.7$&$2.4$ \\
$N_{\rm I\star}$ &{$48.5$}&$48.5$&$48.5$&$48.7$&$47.5$&$47.4$
\\ \hline
$\sgx/M$ & $6.02$&$5.9$&$5.97$&$16.8$&$1.58$&$1.565$ \\
$\sigma_{\rm f}/M$&$1.43$&$1.45$&$1.42$&$2$&$1.41$&$1.41$ \\
$\bar\sigma_{\rm min}/M$&$21.5$&$18$&$21.3$&$25.7$&$2.4$&$1.9$ \\
$\sigma_{\rm max}/M$&$11.9$&$-$&$10.5$&$-$&$1.8$&$-$ \\
$\sigma_{\rm min}/M$&$30.8$&$-$&$32.4$&$-$&$3.1$&$-$
\\ \hline
$10^{9}V'(\bar\sigma_{\rm min})/\kappa^2 M^3$&$-136$&$2.7$&$-141$
&$855$&$-0.89$&$0.46$ \\
$\Dmx~(\%)$&$49$&$-$&$43$&$-$&$12$&$-$\\
$\Dcx~(\%)$&$76$&$76$&$76$&$91.5$&$10.4$&$9.7$\\\hline
$\ns$ & $0.971$&$0.974$&$0.968$&$0.979$&$0.974$&{$0.974$} \\
$-\as/10^{-4}$ & $3.9$&$ $&$4.3$&$5.4$&$2.9$&$3.8$ \\
$r/10^{-7}$ & $1.2$&$1.4$&$0.8$&$15$&$4.8\times10^{-4}$&$2.9\times10^{-4}$ \\
\hline
$\msn/\ZeV$ & $85.2$&$89$&$77.5$&$277$&$3.8$&$3.4$\\
\hline
\end{tabular}
\end{center}
\caption[]{\sl\small Input and output parameters of six Benchmark
points (BMPs) compatible with Eqs.~(\ref{nhi}), (\ref{prob}) and
(\ref{nstot}) for $\Nr=2$ and $N>0$ with [without] sizable $\cssb$
(A, B [C, D]) or $N<0$ (E, F).}\label{tab}
\end{table}

\renewcommand{\arraystretch}{1.}

Comparing the plots of \Fref{fig2} with different $\Nr$ values  we
can appreciate its impact on our results. Namely, we employ
respectively the $\Nr$ values shown in Eq. (\ref{gbl}),
(\ref{glr}) and (\ref{gfl}) in the upper, middle and lower row of
the plots of \Fref{fig2}. We observe that the allowed $\kp$ and
$N$ values are similar in all cases but the $M$ values increase
with $\Nr$ without, though, to approach the value favored by
\Eref{Mgut}. On the other hand, the $\kp-N$ allowed region is
slightly reduced for larger $\Nr$ values. Along the dark blue
curves $\Dmx$ in \Eref{dms} varies in the following ranges
$(0.37-0.66)$ for $\Ggut=\Gbl$, $(0.36-0.65)$ for $\Ggut=\Glr$ and
$(0.34-0.63)$ for $\Ggut=\Gfl$. Therefore, even in the case of the
non-monotonic $\Vhi$ the tuning in the initial conditions is very
low and similar to that needed for quasi-canonical $K$'s -- cf.
\cref{hinova}.

\begin{figure}[!t]\vspace*{-.25in}
\begin{minipage}{0.48\textwidth}
  \includegraphics[width=\linewidth]{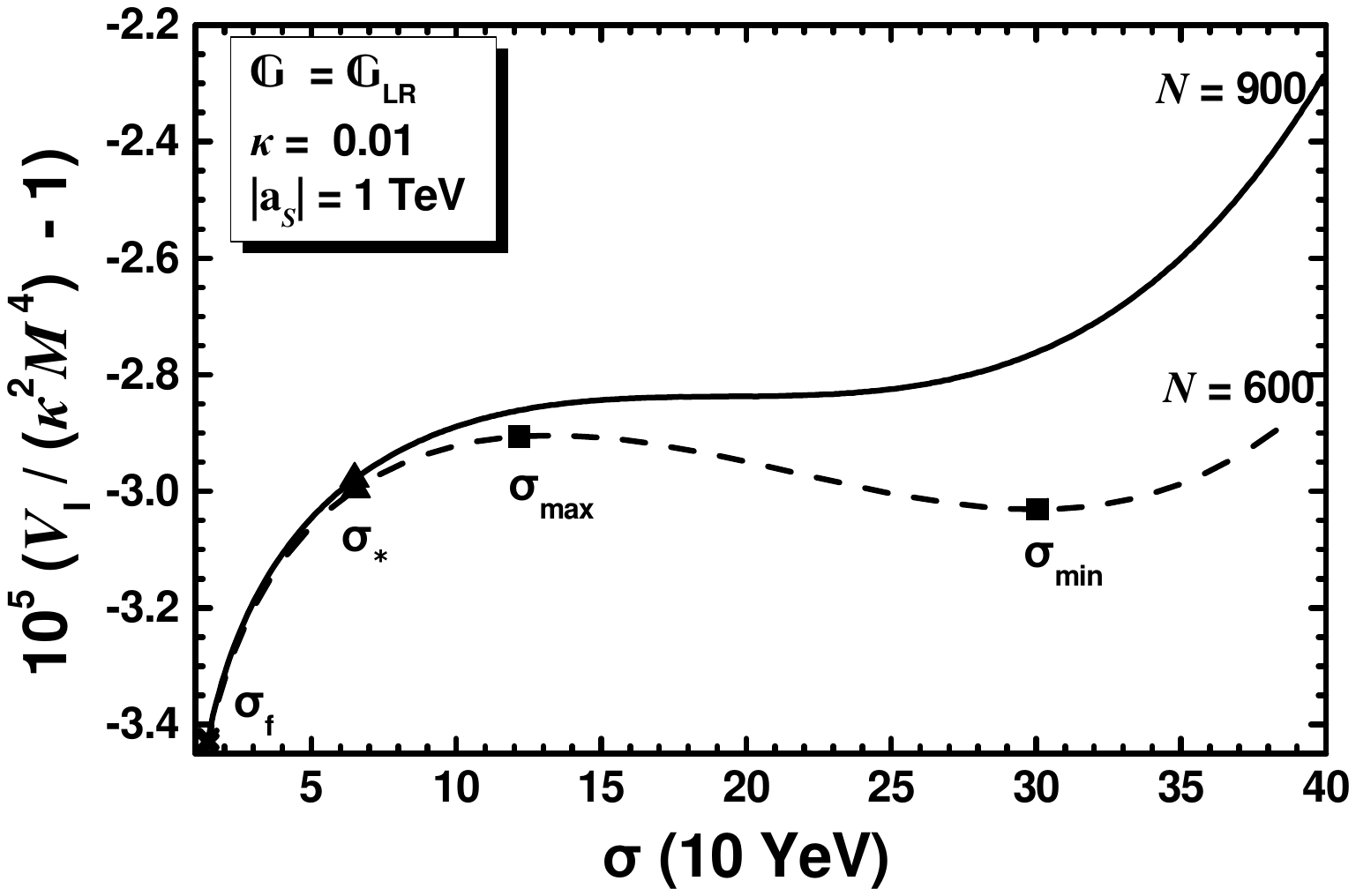}
\end{minipage}\hfill
\begin{minipage}{0.48\textwidth}
    \includegraphics[width=\linewidth]{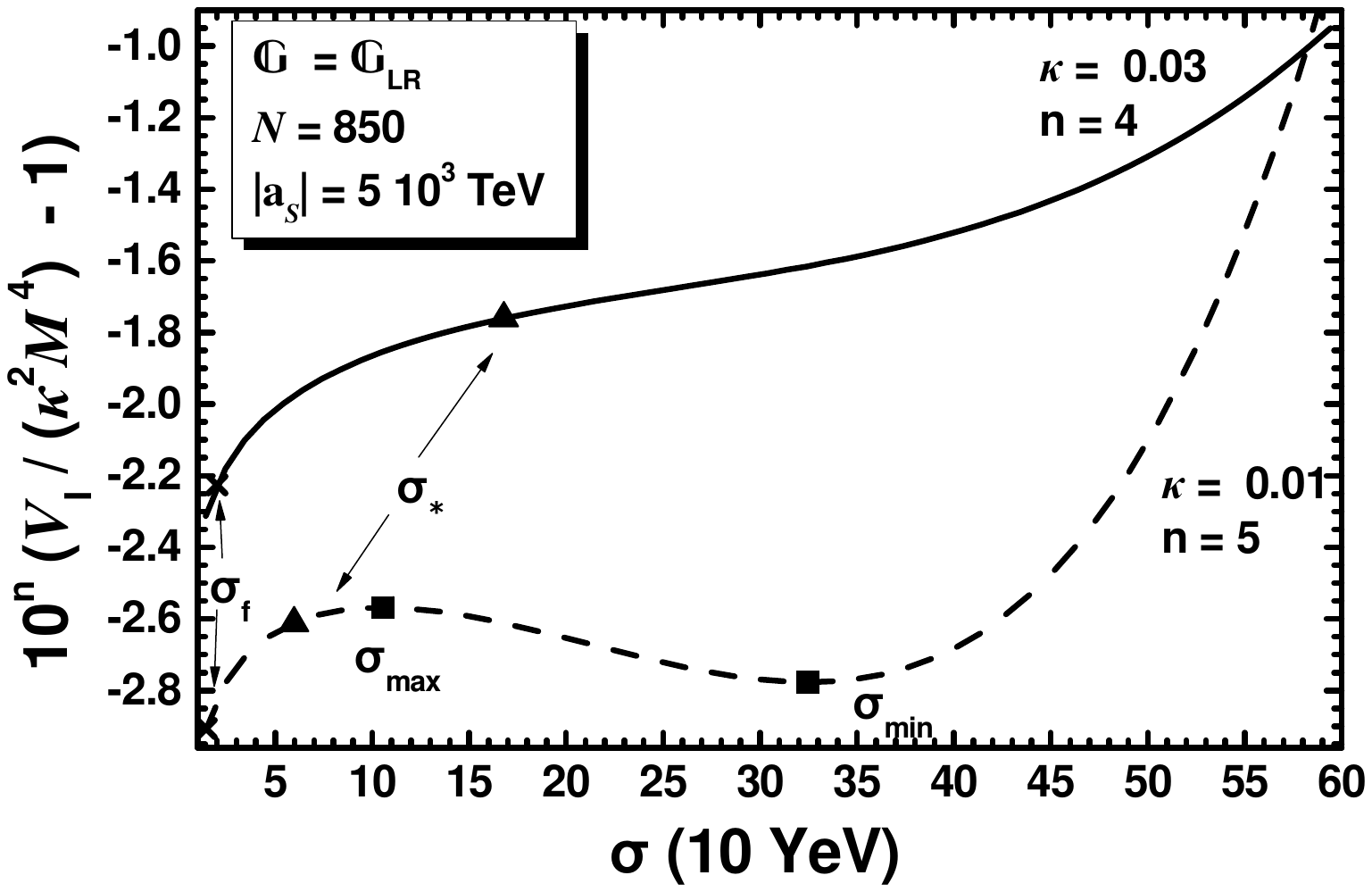}
\end{minipage}\\[-1cm] \centering\includegraphics[width=0.48\linewidth]{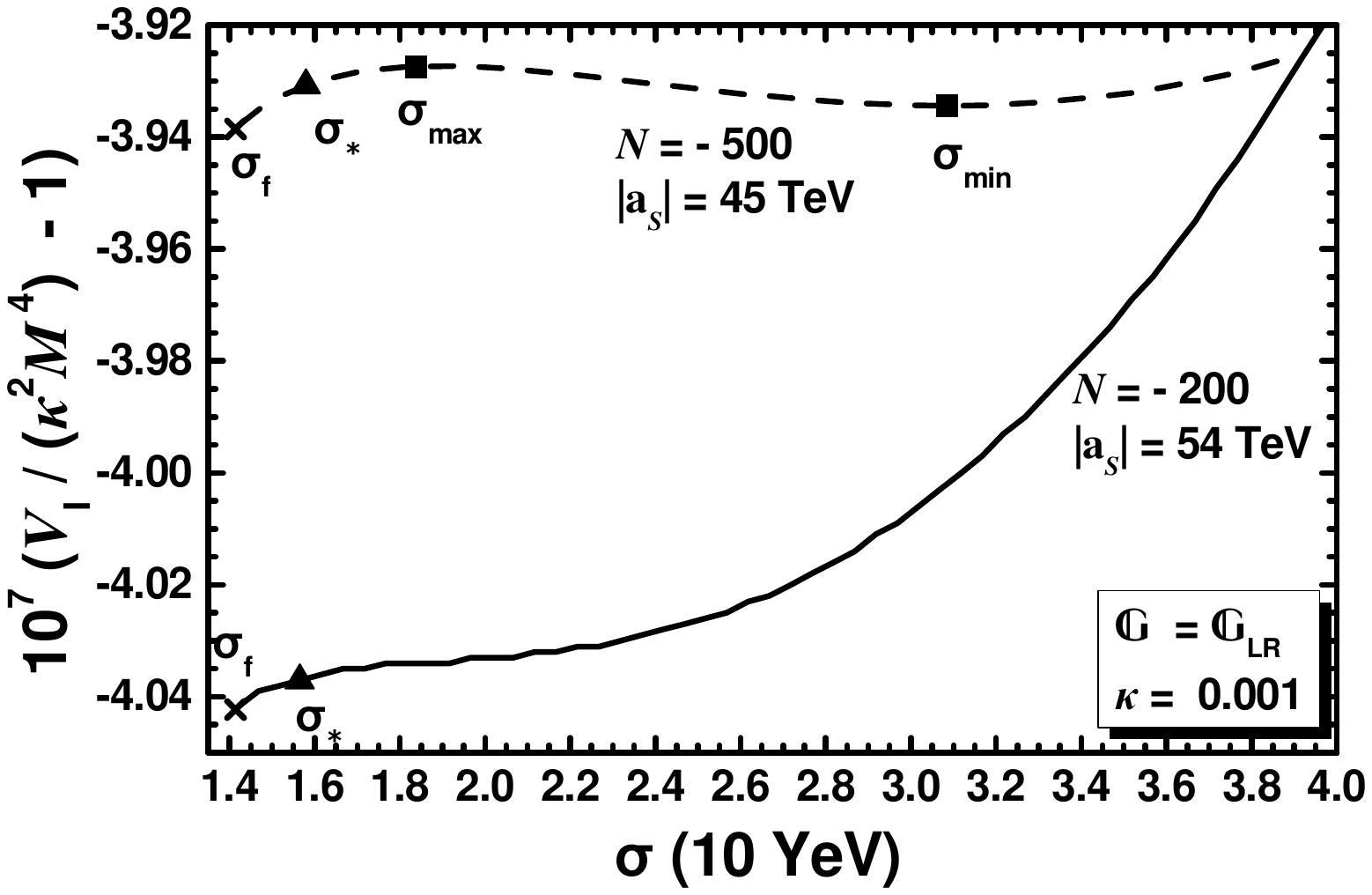}
\vspace*{-.0in} \hfill \caption[]{\sl\small Variation of $\Vhi$ in
\Eref{vhi} as a function of $\sg$ for $\Ggut=\Glr$ and the other
inputs ($\kappa, N$ and $\aS$) of the BMPs in \Tref{tab}. The
values of $\sgx, \sgf, \sgm$ and $\sg_{\rm min}$ are also
depicted. }\label{fig1}
\end{figure}

To gain further insight into the behavior of the solutions in the
green and white regions of the plots of \Fref{fig2} we list in
\Tref{tab} two representative \emph{Benchmark points} ({\sf\ftn
BMPs}) A and B. In both \bmps we fix $\kappa=0.01$, $\Nr=2$ and
$\aaS=1~\TeV$. In \bmp A, however, we select $N=600$ and we obtain
$\Vhi'<0$ -- see \Eref{vmon} -- indicating the development of a
local minimum and maximum of $\Vhi$ whereas in \bmp B we use
$N=820$ yielding $\Vhi'>0$, which signals the monotonic structure
of $\Vhi$. The variation of $\Vhi$ for the \bmps A and B is
plotted in the left panel of the upper row of \Fref{fig1} where
the aforementioned results regarding the monotonicity of $\Vhi$
are verified. In the same plot we also depict the points $\sgx$
and $\sgf$ for the \bmps A and B and $\sgm$ and $\sgn$ for \bmp B.
From the analytic expressions in \eqss{smn0}{nhian}{nsan} we
obtain $\sgm\simeq11.1M$, $\sgn\simeq30.7M$, and $\ns\simeq0.97$
for \bmp A and $\ns\simeq0.973$ for \bmp B in good agreement with
the numerical findings listed in \Tref{tab}. From that Table
we can also infer that the inflaton mass $\msn$ -- defined below
\Eref{vevs} -- lies in the $\ZeV$ range whereas the quantities in
\Eref{dms} do not signal any disturbing tuning.

\begin{figure}[t]
    \centering
     \begin{subfigure}[b]{0.45\textwidth}
        \centering
        \includegraphics[width=\textwidth]{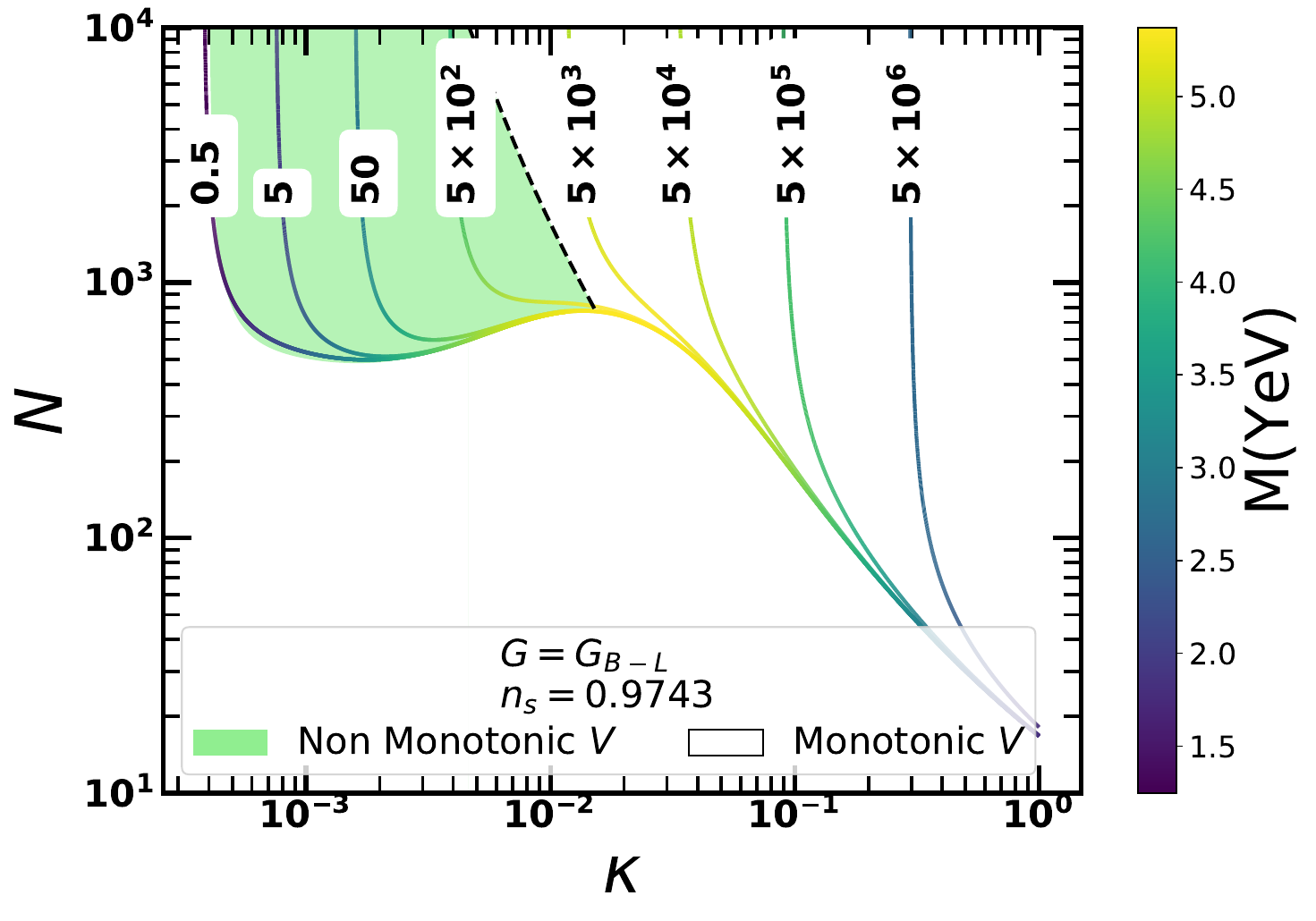}
            \end{subfigure}
            \hspace{0.0091\textwidth} 
    \begin{subfigure}[b]{0.45\textwidth}
        \centering
        \includegraphics[width=\textwidth]{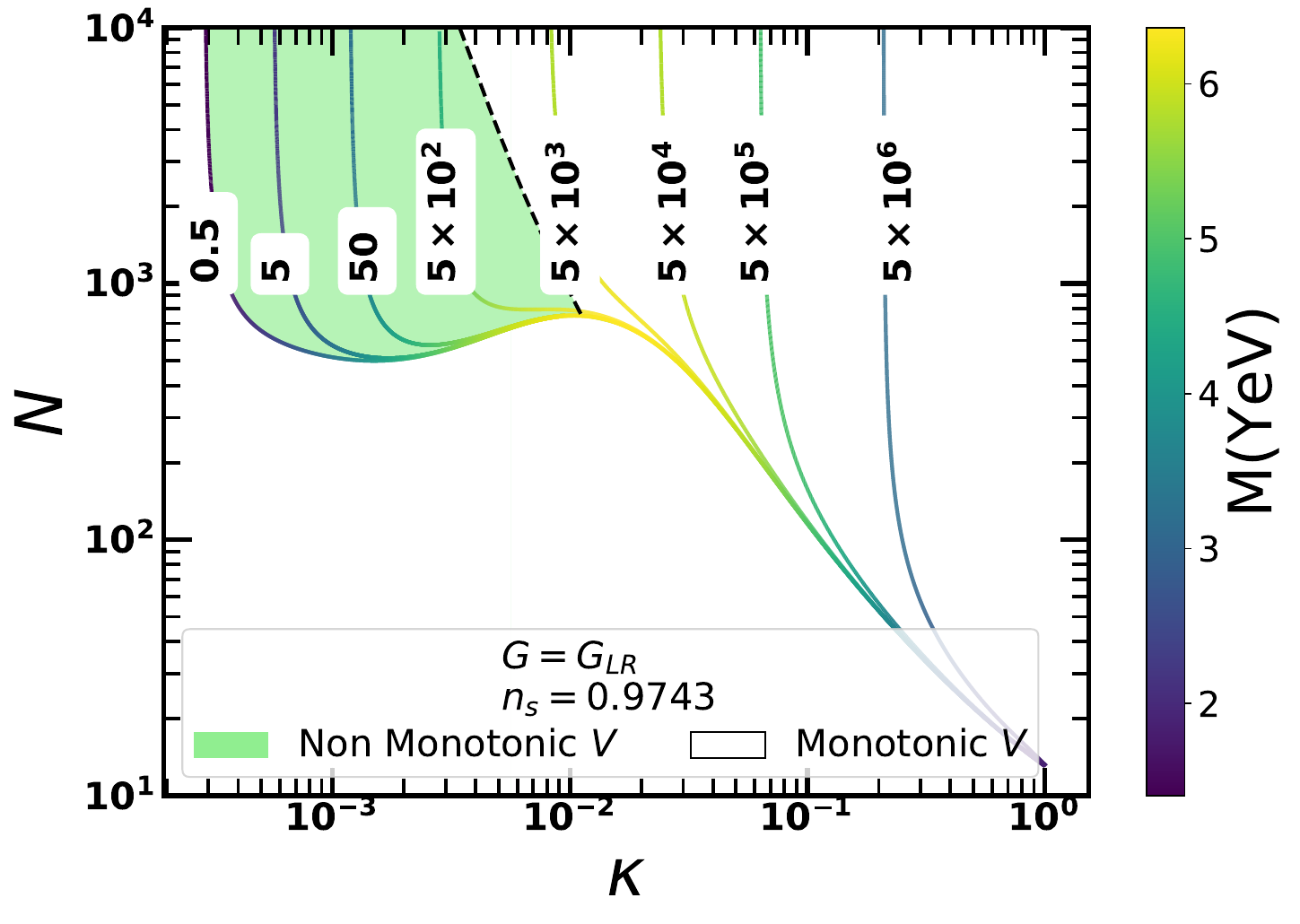}
    \end{subfigure}
            \hspace{0.0091\textwidth} 
    \begin{subfigure}[b]{0.45\textwidth}
        \centering
        \includegraphics[width=\textwidth]{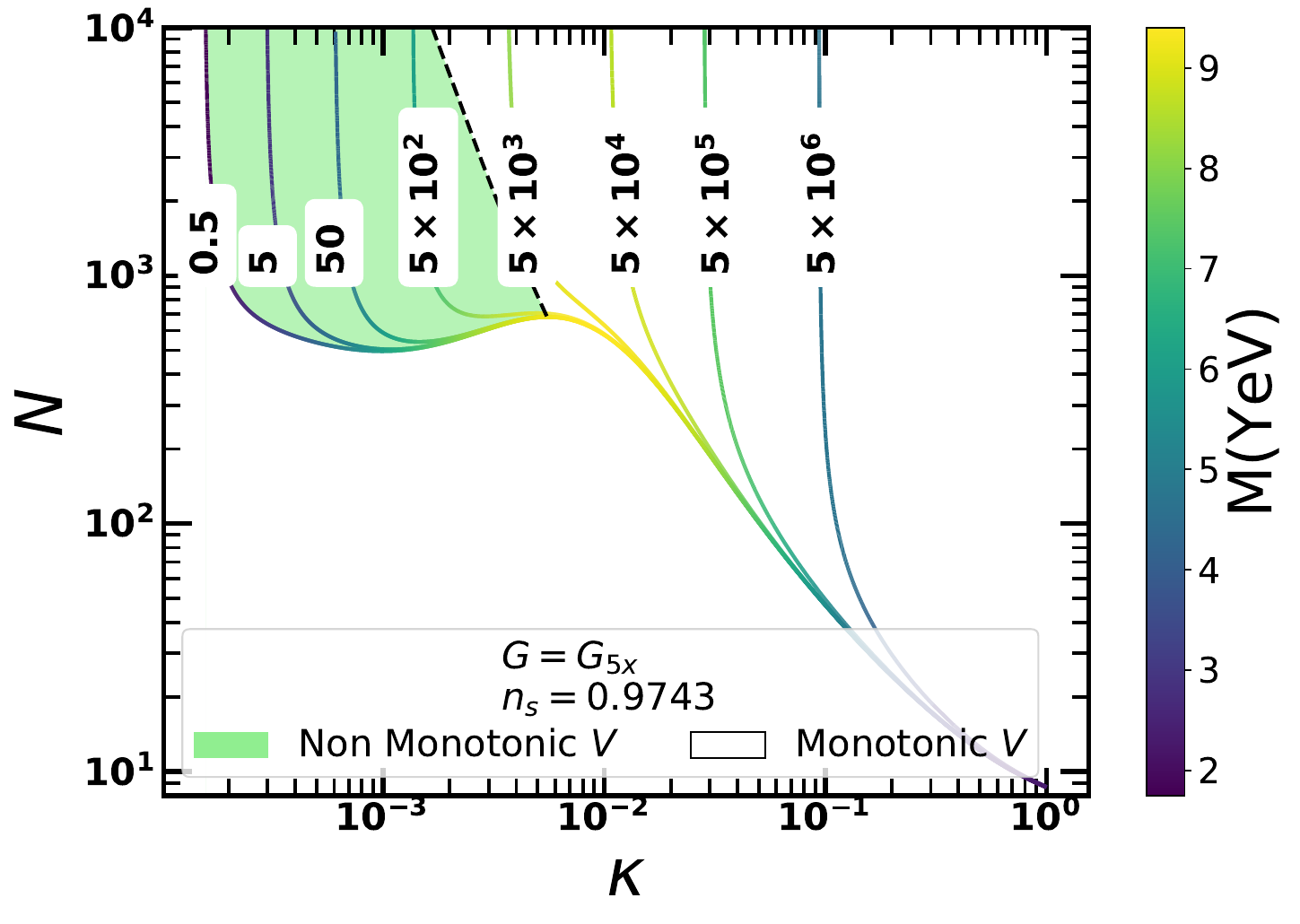}
    \end{subfigure}
\caption[]{\sl\small Curves allowed by \eqs{nhi}{prob} in the
$\kp-N$ plane for central $\ns$ in \Eref{nsact} and various
$|\aS|/\TeV$ and $\Ggut$'s shown on the curves and in the legends
respectively. The color coding of the curves indicates the
variation of $M$. The curves within the green shaded region
correspond to non-monotonic $\Vhi$. } \label{fig3}
\end{figure}

\subparagraph{\sf{\ftn (b)} $N>0$ {\ftn With} $\aaS$ {\ftn
Dependence}.} Related results are presented in \Fref{fig3}, where
we depict curves allowed by \eqs{nhi}{prob} in the $\kp-N$ plane
for central $\ns$ in \Eref{nsact}. We employ various $|\aS|/\TeV$
values shown on the curves and the three $\Ggut$'s indicated in
legends of the plots. We focus on the central $\ns$ value in
\Eref{nsact} since the results for other $\ns$ values in
\Eref{nstot} are similar as we show in \Fref{fig2}. The color
coding of the curves indicates the variation of $M$. As in
\Fref{fig2} we observe that $M$ increases with $\Nr$ but it turns
out to be lower than its values in \Fref{fig2} for fixed $\Nr$.
Therefore, it remains lower than its value required by \Eref{Mgut}
too.

The line for $\aaS/\TeV=0.5$ is identical to the dark blue line
included in \Fref{fig2}. Also, for low $\aaS$ values the various
lines share the same segment with the dark blue line in the
corresponding plots of \Fref{fig2} and for a sizable range of
$\kp$ values. Increasing $\aaS$, however, the almost vertical part
of the allowed curves is shifted to the right and the common part
of the lines with that in \Fref{fig2} is reduced. Interestingly,
for $\aaS>5~\PeV$ the obtained lines lie outside the green shaded
region assuring the monotonicity of $\Vhi$. On the other hand, the
curves within the green shaded region correspond to non-monotonic
$\Vhi$. The transition between these regimes is illustrated by a
black dashed line derived from the condition in \Eref{vmon}. We
observe that for $\kappa \gtrsim 0.01$, monotonic solutions are
preferred, particularly for $\aaS$ in the range $(1-10^{3})~\PeV$.
This indicates that, for moderately large couplings, FHI favors
$\Vhi$ stabilized by substantial soft SUSY-breaking contributions,
consistent with either split \cite{split} or high-scale
\cite{high} SUSY scenarios.

\begin{figure}[t]
    \centering
    \begin{subfigure}[b]{0.45\textwidth}
    \centering
    \includegraphics[width=\textwidth]{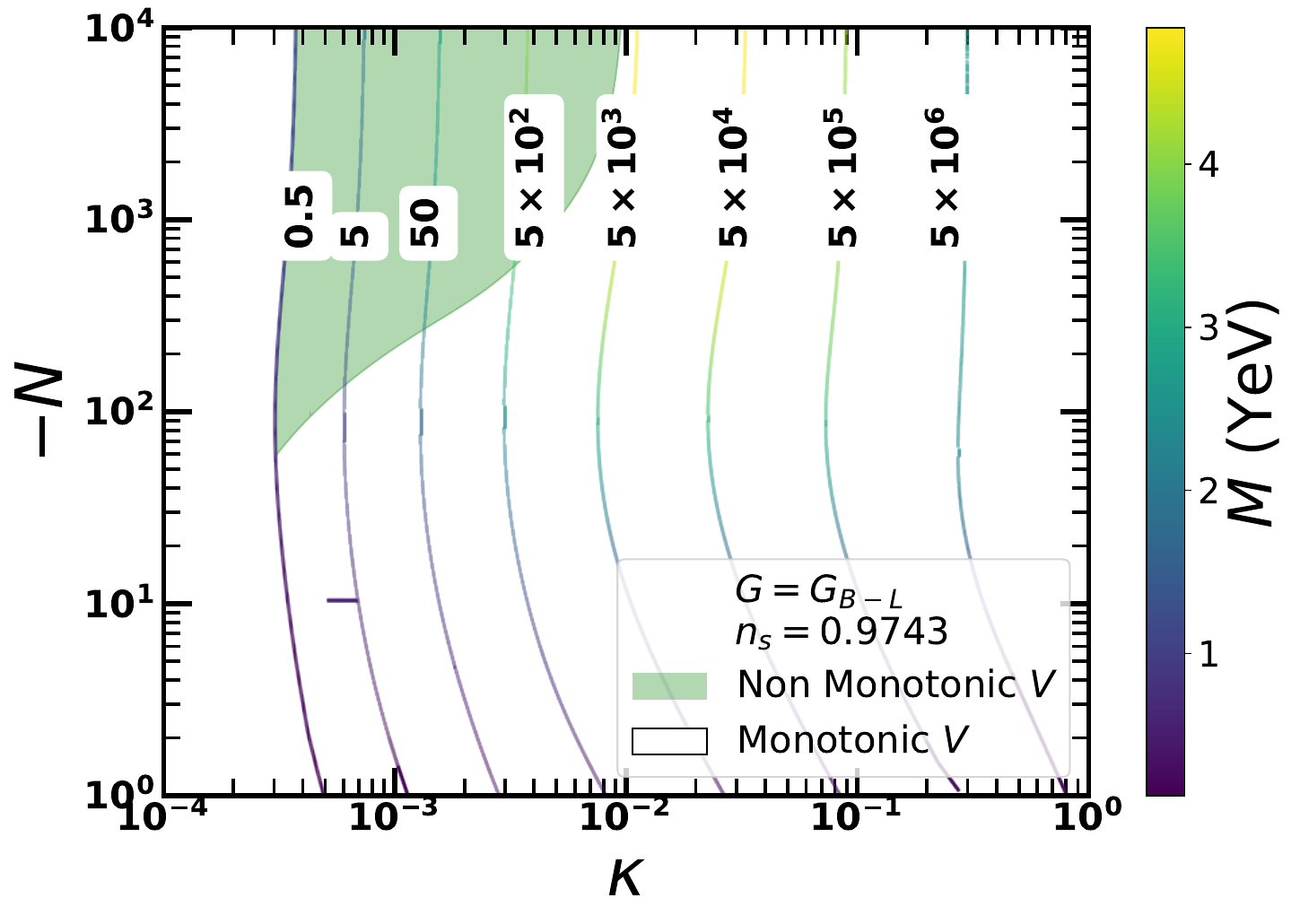}
    \end{subfigure}
    \hspace{0.0091\textwidth} 
    \begin{subfigure}[b]{0.45\textwidth}
    \centering
    \includegraphics[width=\textwidth]{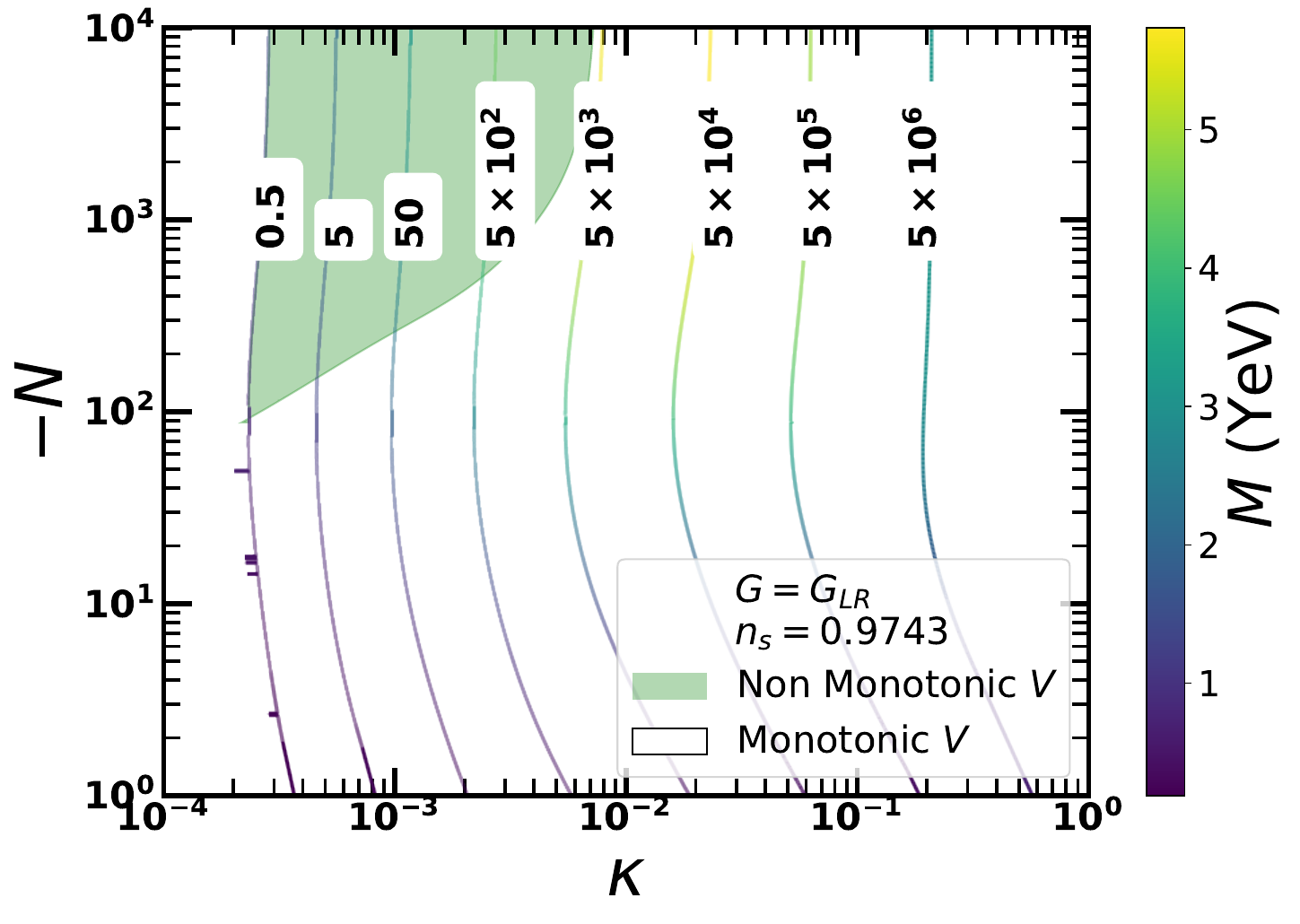}
    \end{subfigure}
      \hspace{0.0091\textwidth} 
    \begin{subfigure}[b]{0.45\textwidth}
    \centering
    \includegraphics[width=\textwidth]{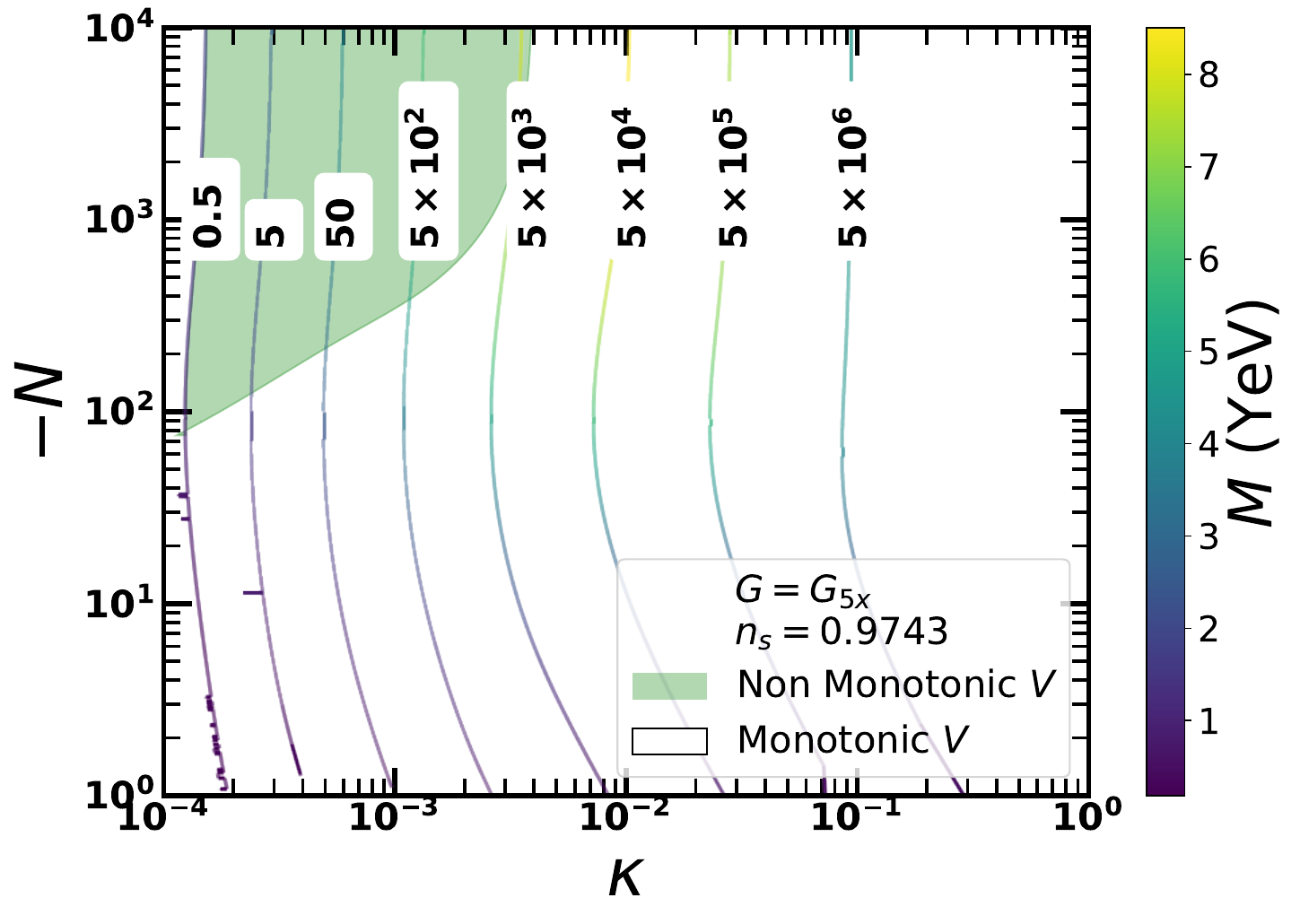}
    \end{subfigure}
\caption{\sl\small Same as Fig.~\ref{fig3} but for $N<0$. }
\label{fig4}
\end{figure}

The salient features of our inflationary solutions in the green
and white regions of the plots of \Fref{fig3} can be further
appreciated by \bmps C and D in \Tref{tab}. In both BMPs we fix
$N=850$, $\Nr=2$ and $\aaS=5~\PeV$. In \bmp C, however,
$\kappa=0.01$ and so $\Vhi'<0$ -- see \Eref{vmon} -- {resulting in
non-monotonic} $\Vhi$ whereas in \bmp D $\kp=0.03$ yields
$\Vhi'>0$ assuring the monotonicity of $\Vhi$. The variation of
$\Vhi$ for the \bmps C and D is plotted in the rightmost panel of
the upper row of \Fref{fig1} where the aforementioned results
regarding the monotonicity of $\Vhi$ are verified. Due to
different $\kappa$ and $M$ the two curves differ appreciably and
for this reason we use different normalization for the variation
of $\Vhi$. In the same plot we also depict the points $\sgx$ and
$\sgf$ for the \bmps C and D and $\sgm$ and $\sgn$ for \bmp C.
From the analytic expression in \Eref{smp} we obtain $\sgm=10M$ in
accordance with the numerical finding listed in \Tref{tab}. From
the same Table we also remark that no tuning issue regarding
$\Dcx$ arises -- despite the fact that $\aaS$ is now sizable --
whereas $\Dmx$ for \bmp C is similar to that for \bmp A.

\subparagraph{\sf{\ftn (c)} $N<0$.} Our results are presented in
\Fref{fig4} where we illustrate curves compatible with
\eqs{nhi}{prob} in the $\kp-N$ plane for central $\ns$ in
\Eref{nsact}. We use the same $|\aS|/\TeV$ values -- shown on the
curves -- as in \Fref{fig3}. Also the {three} panels correspond to
the three $\Ggut$'s considered and specified in the legends of
each panel. As in the previous figures, within the white
region $\Vhi$ enjoys monotonicity whereas within the green shaded
region $\Vhi$ turns out to be non-monotonic. The color coding
along the curves indicates the corresponding values of $M$ which
increase with $\Nr$ but remain lower than the corresponding values
in Fig.~\ref{fig2} and \ref{fig3}. We remark that now each $\kp$
value corresponds, essentially, to one specific $\aaS$ value
almost independently from the $N$ value. This fact signals that
the mechanism described in \cref{mfhi, kaihi} is activated
to obtain the correct $\ns$ value. Contrary to those papers,
however, here we find regions with monotonic $\Vhi$ thanks to the
presence of the quadratic part of $\csgr$ in \Eref{csgr}.

To complete our investigation for $N<0$ {we collect two
representative} \bmps E and F in \Tref{tab}. In both \bmps we fix
$\Nr=2$ and $\kappa=0.001$. In \bmp E we set $|N|=500$ and
$\aaS=45~\TeV$ which yield a non-monotonic $\Vhi$ since $\Vhi'<0$
-- see \Eref{vmon} -- whereas in \bmp F we select $|N|=200$ and
$\aaS=54~\TeV$ leading to monotonic $\Vhi$ since $\Vhi'>0$. The
variation of $\Vhi$ for the \bmps E and F is depicted in the plot
of the lower row of \Fref{fig1} where the aforementioned results
regarding the monotonicity of $\Vhi$ are verified. In the same
plot we also indicate the points $\sgx$ and $\sgf$ for the \bmps E
and F and $\sgm$ and $\sgn$ for \bmp E. From the analytic
expressions in \eqs{sbn}{smn} we obtain $\sgb=2.3M$ [$\sgb=1.6M$]
$\sgm=1.6M$ and $\sgn=3.12M$ for \bmp E [F] in good agreement with
the numerical findings shown in \Tref{tab}. From the displayed
values of $\Dmx$ and $\Dcx$ we notice that $\Dmx$ is lower than
the ones  obtained in \bmps A and C whereas $\Dcx$ is much lower
than those obtained for $N>0$. Therefore, this case can be
characterized, generally, as less natural regarding the aspect of
tuning.

\subparagraph{} The predicted values of $|\as|$ and $r$ within our
scheme remain very small in all cases analyzed above. In
particular, $|\as| \sim 10^{-4}$ remains essentially unchanged
across the range of $\aaS$. In contrast, $r$ increases noticeably
for large $\aaS$, reflecting the growing influence of $\cssb$.
{For $N>0$, $r$ rises up to} $\sim 2.3 \times 10^{-5}$ whereas for
$N<0$, its largest value is $3.3 \times 10^{-6}$ for large $\aaS$
values. These results demonstrate that, while $N$ and $\aaS$ can
affect $r$, it remains even at its maximum well below the
sensitivity of current and near-future CMB and gravitational wave
experiments.

\section{FHI and Cosmic Strings}\label{cssec}

When $\Ggut=\Gbl$, {type I -- according the terminology of
\cref{wells, afzal} --} CSs may be produced after FHI with
dimensionless tension \cite{mfhi}
\begin{equation} \label{mucs} \mcs \simeq
\frac12\lf\frac{M}{\mP}\rg^2\ecs(\rcs)~~\mbox{with}~~\ecs(\rcs)=\frac{2.4}{\ln(2/\rcs)}~~
\mbox{and}~~\rcs=\frac{\kappa^2}{8g^2}\leq10^{-2},\end{equation}
where $G=1/8\pi\mP^2$ is the Newton gravitational constant and
$g\simeq0.7$ is the gauge coupling constant at a scale close to
$M$. {We also take into account} that $(B-L)(\phc)=2$ in
accordance with our assumptions in \Tref{tab1}. The $B-L$ charge
of $\phc$ is determined by the superpotential coupling between
$\phcb$ and the right-handed neutrinos, $\sni$. We here assume a
renormalizable coupling of the form $\phcb(\sni)^2$ as in
\cref{mfhi} -- cf. \cref{actfhi1}. {The resulting uncertainty in
$\mcs$ may induce} minor changes to our results.

If the CSs are \emph{stable}, the corresponding parameter space is
fully allowed by the level of the CS contribution to the observed
anisotropies of CMB which is {constrained} by \plk\ \cite{plcs},
using field-theory simulations of the Abelian-Higgs action -- cf.
\cref{moss} --, in the range
\beq \mcs\lesssim 2.4\times 10^{-7}~~\mbox{at 95$\%$ c.l.}
\label{plcs} \eeq
On the other hand, the recent PTA bound requires \cite{nano1}
\beq \mcs\lesssim 2\times 10^{-10}~~ \mbox{at 95$\%$ c.l.,}
\label{ppta} \eeq
{which excludes the $\mcs$ values predicted within our model} --
see below. However, if these CSs are \emph{metastable} due to the
embedding of $\Gbl$ into a larger gauge group -- such as the
Pati-Salam \cref{nasri}, the flipped $SU(5)$ \cite{leont} or
$SO(10)$ \cite{so10, so10a, so10b, so10c} --, whose spontaneous
breaking to $\Gbl$ produces monopoles, an explanation of the PTA
data on the GWs is possible if $\mcs$ { lies} within the range
preferred by the interpretation of the \emph{NANOGrav 15-year
data} ({\sf\ftn \nano}) \cite{nano1}
\beq  4.3\times10^{-8}\lesssim  \mcs\lesssim 2.4\times
10^{-4}~~\mbox{for}~~8.2\gtrsim\sqrt{\rms}\gtrsim7.5~~ \mbox{at
95$\%$ c.l.}\label{kai} \eeq
Here, the metastability factor $\rms$ is the ratio of the monopole
mass squared, $M_{\rm m}^2$, to $\mu$.  Given that $M_{\rm m}$ is
related to the symmetry breaking scale of {the GUT framework
containing} $\Gbl$, the rightmost restriction in \Eref{kai} may
constrain the relevant scale close to the $M$ shown in \fref{fig5}
-- see below.

The $\mcs$ range in \Eref{kai} can be {more stringently
constrained from above}  from the \emph{LIGO-Virgo-KAGRA}
({\sf\ftn LVK}) collaboration, which implies \cite{ligo}
\beq \mcs \lesssim  2 \times 10^{-7}~~ \mbox{at 95$\%$
c.l.}\label{lvk} \eeq
Note that larger values of $\mcs$ are fully compatible within
nonstandard cosmological scenarios, {such as those discussed in}
\cref{blfhi,antus,actfhi1}. {Assuming here, however,} a
conventional post-inflationary evolution, {as assumed in the
derivation} of \Eref{nhi}, we end up with the following preferred
region in the case of metastable CSs
\beq 4.3 \times 10^{-8} \lesssim \mcs \lesssim 2 \times 10^{-7}.
\label{mcsdata}\eeq

Taking as input the $M$ values for $\Ggut = \Gbl$ and $N>0$ in
\Fref{fig3} or $N<0$ in \Fref{fig4} -- which are consistent with
the inflationary requirements in \eqs{nhi}{prob} and assure
$\ns=0.974$ -- we can compute through \Eref{mucs} the predicted
$\mcs$. The results are presented in \Fref{fig5} which shows
$\mcs$ as a function of $N$ for $N>0$ (left panel) and $-N$ for
$N<0$ (right panel). Each curve corresponds to a fixed value of
$|\aS|$, ranging from $0.5$ TeV to $5\times10^{5}~\TeV$, and the
color coding along each curve indicates the variation of $M$ (in
units of $1~\YeV$). In both panels we also include the upper bound
on $\mcs$ in \Eref{plcs}, marked by a horizontal red dashed line,
and the margin of \Eref{mcsdata} depicted by the light red shaded
region. {We note that} most of the parameter space in
Fig.~\ref{fig4} lies above the much stricter stable-CS bound in
\Eref{ppta}, confirming that stable CSs are ruled out by this
limit.

\begin{figure}[!t]
    \centering
    \begin{subfigure}[b]{0.48\textwidth}
        \centering
        \includegraphics[width=\textwidth]{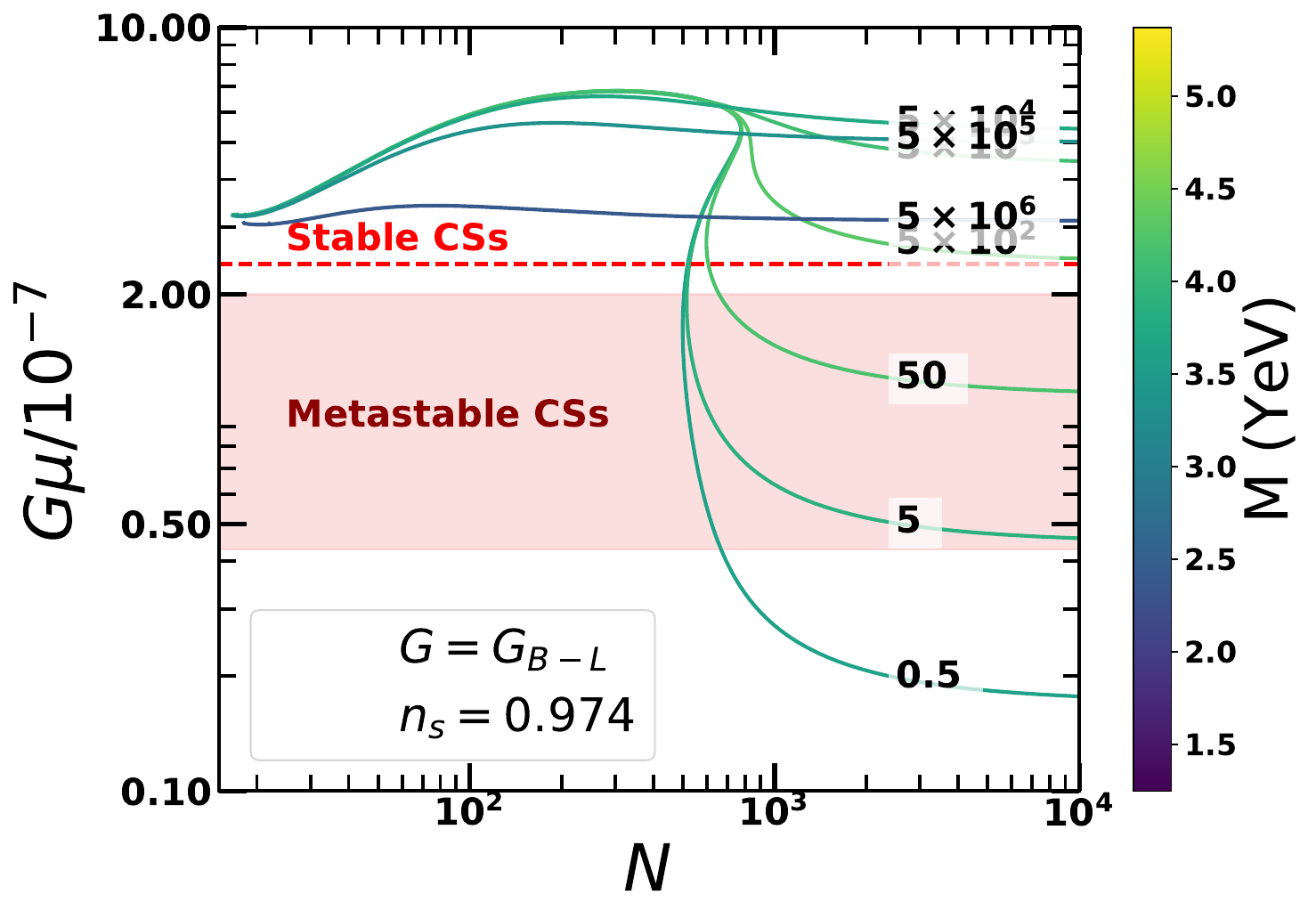}
    \end{subfigure}
    \hspace{0.0091\textwidth} 
    \begin{subfigure}[b]{0.47\textwidth}
        \centering
        \includegraphics[width=\textwidth]{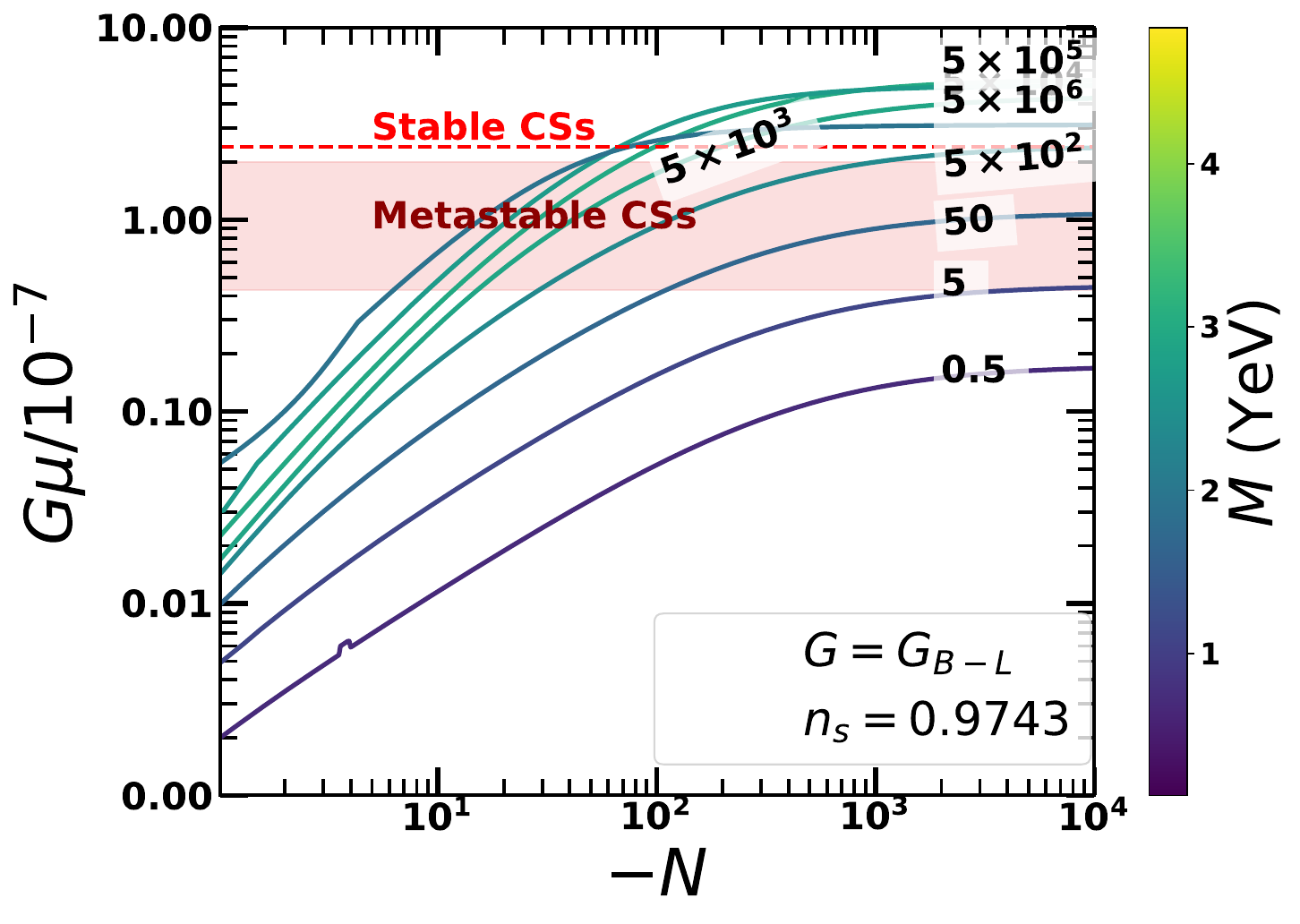}
    \end{subfigure}
\caption{\sl\small $\mcs$ as a function of $N$ for $N>0$  (left
panel) and $-N$ for $N<0$ (right panel). We use $\ns=0.974$ and
various $|\aS|/\TeV$ values shown on the curves. The color coding
of the curves indicates the variation of $M$. The upper bound on
$\mcs$ in \Eref{plcs} and the margin of \Eref{mcsdata} are also
depicted by the horizontal red dashed line and the light red
shaded region respectively. } \label{fig5}
\end{figure}

From the left panel we observe that $\mcs$ decreases monotonically
as $N$ increases beyond $500$ and therefore we obtain
compatibility with \eqs{plcs}{mcsdata} for low $\aaS$ values,
i.e., for $\aaS\simeq(0.5-500)~\TeV$. On the contrary, from the
right panel we remark that $\mcs$ increases with $|N|$ and so for
the entire displayed range of $N$ we can find $\aaS$ which
{provides coverage of the range} of \Eref{mcsdata} favored by
NG15, i.e., for $5\lesssim\aaS/\TeV\lesssim5\times10^6$. The
consistency with NG15 seems to prefer, for $N>0$, the region in
\Fref{fig3} where $\Vhi$ {becomes non-monotonic}. On the contrary,
for $N<0$, regions from \Fref{fig4} where $\Vhi$ is monotonic can
be compatible with NG15. Thus, Fig.~\ref{fig5} provides a direct
link between the internal geometry of the inflationary sector via
$N$, the soft SUSY-breaking scale via $\aaS$, and the observable
GW background from CSs.

\section{Conclusions}\label{con}

We checked the viability of FHI (i.e., F-term hybrid inflation)
under the assumption that the gauge-singlet inflaton field obeys
hyperbolic or compact \Kaa geometry characterized by a single
curvature parameter {$N>0$ and $N<0$, respectively} -- see
\Eref{khi}. Our main findings can be summarized as follows:

\begin{itemize}

\item The inflationary potential receives three crucial
corrections: {SUGRA corrections depending on $N$}, soft
SUSY-breaking parameterized by $\aaS$, and radiative corrections
with some {dependence on the dimensionality} $\Nr$ of the
representations of the waterfall fields.

\item Using the latest CMB data we determined the allowed ranges
of the $W$ coupling $\kappa$ and scale $M$, as a function of $N$
for given $\Nr$ and $\aS$ in Figs.~\ref{fig2} and~\ref{fig3} for
$N>0$ and in \Fref{fig4} for $N<0$. {For all three representative
GUT scenarios} ($\Gbl$, $\Glr$, $\Gfl$) and for $N>0$ the model
{admits a wide and natural parameter range} of parameters which
ensure monotonic $\Vhi$. Most notably, for $\ns\simeq 0.974$, we
found
\beq 0.008\lesssim\kp\lesssim0.1,~10\lesssim
N\lesssim850,~5\lesssim
\aaS/\PeV\lesssim5\times10^3~~\mbox{and}~~1\lesssim
M/\YeV\lesssim9,\label{res}\eeq
where $M$ is slightly below than the MSSM unification scale. For
$N<0$,  we also obtain a broad range of parameters with monotonic
$\Vhi$, almost independent of $N$, which suffers, however, from
some tuning associated with the proximity between $\sgx$ and
$\sgc$ -- see \Eref{dms}. Moreover, our acceptable solutions
require a specific correspondence between $\aaS$ and $\kp$ (or
$M$). For example, for $\aaS=5~\PeV$ we find
\beq 0.79\lesssim\kp/10^{-2}\lesssim1.9,~1\lesssim
-N\lesssim10^4~~\mbox{and}~~0.3\lesssim
M/\YeV\lesssim5.65.\label{resn}\eeq
For both signs of $N$ the primordial inflationary gravitational
waves are essentially undetectable by current and future
experiments since $r$ remains below about $10^{-5}$.

\item The effective field theory  remains valid up to the Planck
scale $\mP$, {without issues related to perturbative unitarity} or
higher-order corrections.

\item When $\Ggut = \Gbl$, CSs (i.e. cosmic strings) are produced
at the end of FHI with dimensionless tension in the range $\mcs
\sim 10^{-8} - 10^{-7}$. If CSs are stable, the model is ruled out
by the NG15 upper bound. On the other hand, if the CSs are
metastable the same $\mcs$ values fall within the window that
explains the observed stochastic background of GWs, as shown in
Fig.~\ref{fig5}. Thus, our model provides a simultaneous
explanation of CMB data and the NG15 signal without fine tuning.

\end{itemize}

{Finally, we note} that a complete inflationary scenario must also
account for the transition to a radiation-dominated universe and
the generation of the observed baryon asymmetry. These aspects
have been extensively explored in the context of \fhi\ with
canonical or quasi-canonical \Ka s\ -- see, e.g.,
Refs.~\cite{mfhi,nmfhi}. Our setup preserves many of the
successful features of this post-inflationary evolution, which may
impose additional constraints on the viable parameter space and
help identify the most compelling version of \fhi.


\acknowledgments{{We would like to thank A. Afzal and R. Maji for
interesting comments}. C.P. would like to thank Q. Shafi for
useful discussions and M. Ashry for collaboration during an early
stage of this work.} M. U. R. extends his appreciation to the
Deanship of Scientific Research, Islamic University of Madinah,
Saudi Arabia, for funding this research work.


\newcommand\jcap[3]{{\it JCAP }{\bf #1}, #3 (#2)}
\renewcommand\jhep[3]{{\it JHEP }{\bf
#1}, #3 (#2)}
\newcommand{\arxiv}[1]{{\tt arXiv:#1}}
\def\prd#1#2#3{{\sl Phys. Rev. D }{\bf #1}, #3 (#2)}
\def\prdn#1#2#3#4{{\sl Phys. Rev. D }{\bf #1}, no~#4, #3 (#2)}

\end{document}